\documentclass[a4paper,fleqn,usenatbib, useAMS]{mnras}
\usepackage{pdflscape}	
\usepackage{txfonts}
\usepackage{natbib}
\usepackage{graphicx}
\usepackage{xspace}
\usepackage{longtable}

\usepackage[T1]{fontenc}
\usepackage{ae,aecompl}

\title[Spectropolarimetry of polarimetric standard stars]{Linear spectropolarimetry of polarimetric standard stars with VLT/FORS2\thanks{Based on archival calibration data collected at the European Organisation for Astronomical Research in the Southern Hemisphere.}}

\author[A. Cikota et al.]{
Aleksandar Cikota$^{1}$\thanks{E-mail: acikota@eso.org},
Ferdinando Patat$^{1}$,
Stefan Cikota$^{2}$,
Tamar Faran$^{1}$
\\
$^{1}$European Southern Observatory, Karl-Schwarzschild-Str. 2, 85748 Garching b. M\"{u}nchen, Germany \\
$^{2}$Physics Department, University of Split, Ru\dj era Bo\v{s}kovi\'{c}a 33, 21000 Split, Croatia 
}

\date{Accepted XXX. Received YYY; in original form ZZZ}

\pubyear{2016}

\begin{document}
\label{firstpage}
\pagerange{\pageref{firstpage}--\pageref{lastpage}}
\maketitle

\begin{abstract}
We reduced ESO's archival linear spectropolarimetry data (4000-9000\AA) of 6 highly polarized and 8 unpolarized standard stars observed between 2010 and 2016, for a total of 70 epochs, with the FOcal Reducer and low dispersion Spectrograph (FORS2) mounted at the Very Large Telescope. We provide very accurate standard stars polarization measurements as a function of wavelength, and test the performance of the spectropolarimetric mode (PMOS) of FORS2. We used the unpolarized stars to test the time stability of the PMOS mode, and found a small ($\leq$0.1\%), but statistically significant, on-axis instrumental polarization wavelength dependency, possibly caused by the tilted surfaces of the dispersive element. The polarization degree and angle are found to be stable at the level of $\leq$0.1\% and $\leq$0.2 degrees, respectively. We derived the polarization wavelength dependence of the polarized standard stars and found that, in general, the results are consistent with those reported in the literature, e.g. \citet{2007ASPC..364..503F} who performed a similar analysis using FORS1 data. The re-calibrated data provide a very accurate set of standards that can be very reliably used for technical and scientific purposes. The analysis of the Serkowski parameters revealed a systematic deviation from the width parameter $K$ reported by \citet{1992ApJ...386..562W}. This is most likely explained by incorrect effective wavelengths adopted in that study for the \textit{R} and \textit{I} bands.
\end{abstract}

\begin{keywords}
stars: general -- instrumentation: polarimeters -- (ISM:) dust, extinction
\end{keywords}


\section{Introduction}

Spectropolarimetry is a technique that provides additional information to those given by simple intensity measurements.
For instance, linear spectropolarimetry can probe the geometry of Supernova explosions (see \citet{2008ARA&A..46..433W} for a comprehensive review), circumstellar and interstellar dust properties (see e.g. \citet{1975ApJ...196..261S}), and the magnetic fields of galaxies (see e.g. \citet{1996QJRAS..37..297S}, \citet{1996MNRAS.282..252S}).

Given the comparatively low levels of polarization that are typical in this kind of studies, it is important the polarimeter being used is properly calibrated. The main aim of this work is to provide a re-calibration of a sub-set of highly polarized standard stars, which are not too bright to be observed with large telescopes, and can be used to verify the performance of other similar instruments in the southern hemisphere. In addition, related to the fact that ESO's Quality Control Group monitors polarimetric standards since 2009, we test the performance of the FOcal Reducer and low dispersion Spectrograph (FORS2) instrument mounted at the Very Large Telescope (VLT), in terms of stability and accuracy, by means of a thorough analysis of archival calibration data of highly linearly polarized and unpolarized standard stars.

\citet{2007ASPC..364..503F} presented an analysis of standard stars observations obtained between 1999 and 2005 in PMOS and imaging-polarimetry (IPOL) mode with FORS1. They investigated the FORS1 stability in both, PMOS and IPOL mode, and found a small instrumental offset in the Stokes Q parameter which appears in PMOS mode only. This study is meant to further investigate and characterize that offset using FORS2 data.

\section{Instrumental setup and observations}

FORS2 is currently mounted at the Cassegrain focus of the ESO's Antu VLT unit. This focal reducer is equipped with a polarimetric mode, which can be selected by introducing in the light path a Wollaston prism and a rotatable super-achromatic half-wave retarder plate coupled to a grism and/or a filter to perform linear imaging or spectro-polarimetry \citep{1967PASP...79..136A, 1998Msngr..94....1A, FORS2manual}.

For our purposes, we selected eight standard stars with zero polarization, and six highly polarized standard stars extracted from the FORS Standard Fields and Stars list\footnote{$https://www.eso.org/sci/facilities/paranal/instruments/fors/\\tools/FORS\_Std.html$} (listed in Table~\ref{tab1}). All stars from the FORS Standard Fields and Stars list (i.e. our sample), were also analyzed by \citet{2007ASPC..364..503F}. They additionally considered three polarized standard stars (HD 345310, HD 111579 and BD -13 5073), 8 unpolarized stars (WD 2359-434, WD 0310-688, WD 1616-154, WD 1620-391, HD 176425, WD 2007-303, WD 2039-202, WD 2149+021) and HD 64299, which was erroneously suggested as an unpolarized standard star, but is in fact polarized at the 0.1$\%$ level (see e.g. \citet{2007PASP..119.1126M}).

From the archive we selected observations obtained in PMOS mode with grism 300V, without an order separating filter, and with the half-wave retarder plate positioned at angles of 0$^{\circ}$, 22.5$^{\circ}$, 45$^{\circ}$, and 67.5$^{\circ}$. The half-wave retarder plate angle is measured between the acceptance axis of the ordinary beam of the Wollaston prism (which is aligned to the north-south direction) and the fast axis of the retarder plate.
In order to allow a time trend analysis, the stars were selected to have observations at multiple epochs (40 epochs in total of unpolarized, and 30 of polarized stars), spanning 5 years. The stars were placed on the central slit of the PMOS focal mask, very close to the optical axis of the instrument.
The spectrum produced by the grism is split by the Wollaston prism into an ordinary (o) and extraordinary (e) beam, which are separated by a throw of about 22 arcseconds. 

\begin{table*}
\centering
\caption{\label{tab:stars}List of observed standard stars.}
\label{tab1}
\begin{tabular}{lllrllc}
\hline\hline
Name	 	& RA 			& DEC 				& V		& Spec.    &       & No. of\\
	 &	(J2000)	     	&	(J2000)			&(mag)	&type		& Type & Epochs\\
\hline
HD 10038	  & 01 37 18.59	& -40 10 38.46		& 8.1	& A2mA5-F0	& unPol. & 8\\ 
HD 13588	  & 02 11 16.69 	& -46 35 06.17 		& 7.9	& A1m		& unPol. & 4 \\ 
HD 42078  &  06 06 41.04 	& -42 17 55.69   	& 6.2	& Am		& unPol. & 5\\ 
HD 97689   & 11 13 50.75 	& -52 51 21.22 		& 6.8 	& A0m	& unPol. & 12 \\ 
WD 1615-154	 & 16 17 55.26	& -15 35 51.93 		& 13.4 	& DA1.7		& unPol. & 1\\ 
WD 1620-391	 & 16 23 33.84 	& -39 13 46.16 	 	& 11.0 	& DA2		& unPol. & 7\\ 
WD 2039-202	 & 20 42 34.75	& -20 04 35.95 		& 12.4 	& DA2.5		& unPol. & 2\\ 
WD 2149+021	 & 21 52 25.38 	& +02 23 19.54 	 	& 12.7 	& DA2.8		& unPol. & 1\\ 
NGC 2024 1	 & 05 41 37.85	& -01 54 36.5		& 12.2 	& B0.5V 	& Pol. & 7\\
Vela1 95$^{\mathrm{a}}$ & 09 06 00.01	& -47 18 58.2		& 12.1 	& OB+		& Pol. & 11\\ 
Hiltner 652$^{\mathrm{b}}$ & 17 43 19.59	& -28 40 32.76		& 10.8  & B1II-III	& Pol. & 8\\ 
HDE 316232	 & 17 45 43.70	& -29 13 18.15		& 10.4  & O9IV		& Pol. & 1\\ 
BD -14 4922	 & 18 11 58.10	& -14 56 09.01		& 9.73  & O9.5		& Pol. & 2\\ 
BD -12 5133	 & 18 40 01.70	& -12 24 06.92		& 10.4  & B1V		& Pol. & 1\\ 
\hline
\multicolumn{7}{l}{The coordinates, brightness and spectral type were taken from the SIMBAD Astronomical}\\
\multicolumn{7}{l}{Database. Type indicates if the star is polarized (Pol.) or unpolarized (unPol.), and No. of }\\
\multicolumn{7}{l}{Epochs is the number of epochs.}\\
\multicolumn{7}{l}{$^{\mathrm{a}}$ in \citet{2007ASPC..364..503F} designated as Ve 6-23.}\\
\multicolumn{7}{l}{$^{\mathrm{b}}$ in \citet{2007ASPC..364..503F} designated as CD-28 13479.}\\
\end{tabular}
\end{table*}

\section{Data reduction}
\label{datareduction}

After excluding saturated frames, the data were reduced using standard procedures in IRAF. Wavelength calibration was achieved using He-Ne-Ar arc lamp exposure. The typical RMS accuracy is $\sim$ 0.3 \AA. Although it is very difficult to achieve a complete flat fielding correction in the presence of polarization optics, the effects of improper correction were minimized by taking advantage of the redundant number of half-wave positions (see \citet{2006PASP..118..146P}).

Ordinary and extra-ordinary beams were extracted in an unsupervised way using the PyRAF apextract.apall procedure, with a fixed aperture size of 10 pixels. To avoid spectrum tracing problems, the input frames were properly trimmed to exclude the low signal-to-noise at the edges of the spectral range. The final effective wavelength range is 3950-9300 \AA.
We calculated the total flux in \textit{BVRI} passbands by integrating the total flux weighted with Bessel's \textit{BVRI} passband filters, and computed the synthetic broad-band polarization degree and polarization angle corresponding to each passband. We also binned the spectra in 50 $\AA$ bins, in order to obtain a larger signal-to-noise ratio, and calculated the Stokes parameters Q and U, polarization degree $P$, and polarization angle $\theta_P$ as a function of wavelength. The signal-to-noise ratio (SNR) is typically between 500 and 2000 per 50$\AA$ bin, tabulated for each individual epoch at a given wavelength in Table~\ref{tbunpol} and Table~\ref{tbpol} for unpolarized and polarized standard stars respectively.

The Stokes parameters Q and U were derived via Fourier transformation, as described in the FORS2 User Manual:
\begin{equation}
\begin{array}{l}
Q = \frac{2}{N} \sum_{i=0}^{N-1} F(\theta_i)\cos(4\theta_i)  \\ 
U = \frac{2}{N} \sum_{i=0}^{N-1} F(\theta_i)\sin(4\theta_i)
\end{array}
\end{equation}
where $F(\theta_i)$ are the normalized flux differences between the ordinary ($f^o$) and extra-ordinary ($f^e$) beams:
\begin{equation}
\label{eqF}
F(\theta_i) = \frac{f^o (\theta_i) - f^e (\theta_i)}{f^o (\theta_i) + f^e (\theta_i)}
\end{equation}
at different half-wave retarder plate position angles 
$\theta_i = \textit{i} * 22.5^{\circ}$.

Although FORS2 is equipped with a super-achromatic half wave plate, residual retardance chromatism is present. The wavelength dependent retardance offset ($-\Delta\theta(\lambda)$) is tabulated in the FORS2 User Manual \citep{FORS2manual}. The chromatism was corrected through the following rotation of the Stokes parameters \citep{2009PASP..121..993B, 2011A&A...529A..57P}:
\begin{equation}
\begin{array}{l}
Q_{\mathrm{corrected}} = Q \cos 2\Delta\theta(\lambda) - U \sin 2\Delta\theta(\lambda) \\
U_{\mathrm{corrected}} = Q \sin 2\Delta\theta(\lambda) + U \cos 2\Delta\theta(\lambda)
\end{array}
\end{equation}
Hereafter, $Q_{corrected}$ and $U_{corrected}$ are noted as Q and U.

Finally we calculate the polarization (eq. \ref{eqP}): 
\begin{equation}
\label{eqP}
P=\sqrt{Q^2+U^2}
\end{equation}
and the polarization angle (eq. \ref{eqtheta}), 
\begin{equation}
\label{eqtheta}
\theta = \frac{1}{2}\arctan(U/Q) .
\end{equation}
The statistical uncertainties were estimated using the prescriptions presented by \citet{2006PASP..118..146P}.

The classical Serkowski parameters (\citet{1975ApJ...196..261S}, eq.~\ref{eq_serkowski}) were finally derived by linear least squares fits to the linear polarization $P(\lambda)$, yielding the peak polarization level ($P_{\mathrm{max}}$) and wavelength ($\lambda_{\mathrm{max}}$), and the width constant ($K$).
\begin{equation}
\label{eq_serkowski}
\frac{P(\lambda)}{P_{\mathrm{max}}} = \exp \left[-K \mathrm{ln}^2 \left( \frac{\lambda_{\mathrm{max} }}{\lambda} \right) \right] 
\end{equation}

\section{Results and Discussion}

\subsection{Unpolarized stars}

\begin{figure}
\begin{center}
\includegraphics[trim=20mm 10mm 10mm 10mm, width=9cm, clip=true]{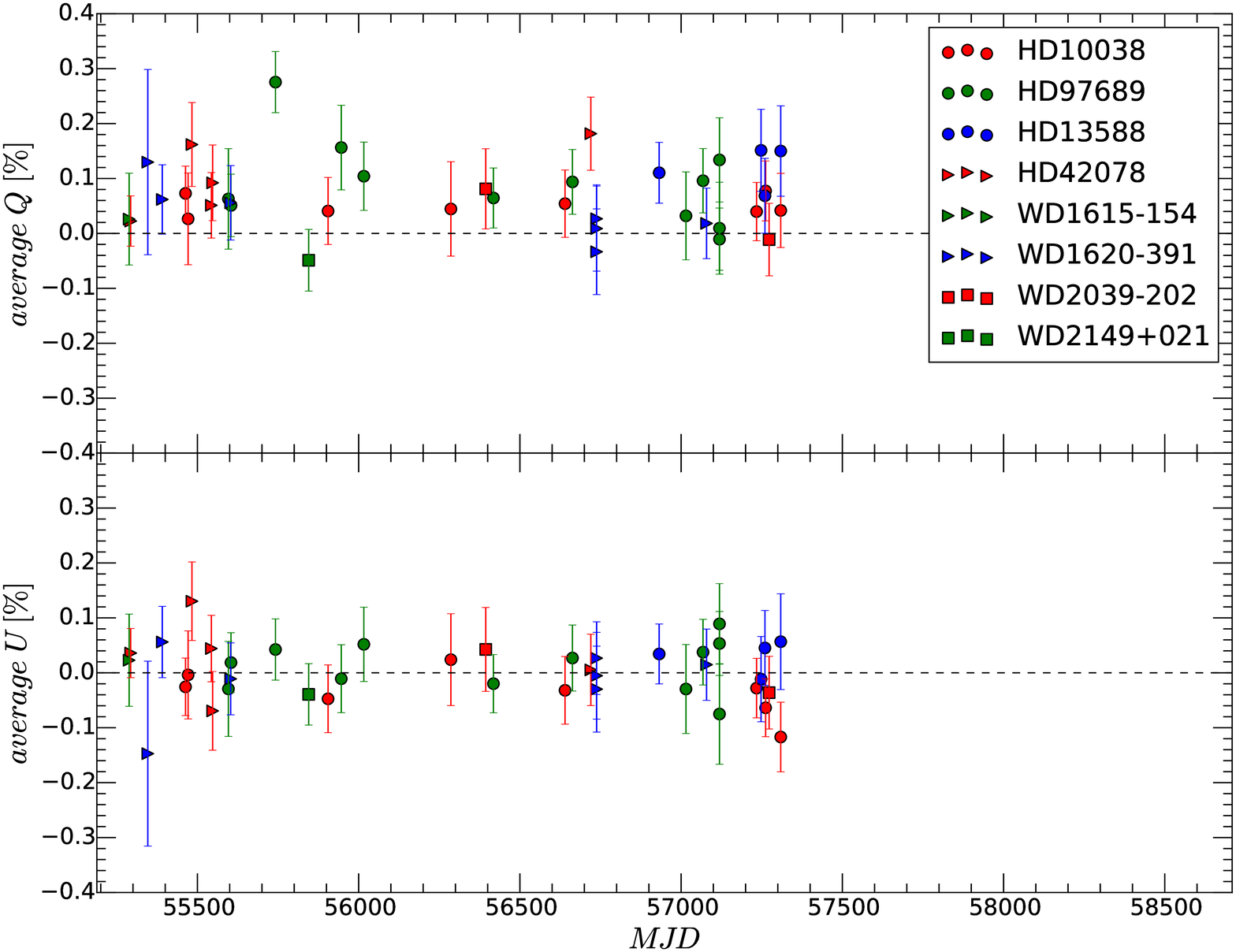}
\vspace{-4mm}
\caption{Averaged Stokes parameters $P_Q$ and $P_U$ for the 8 unpolarized standard stars in the sample observed on 40 different epochs.}
\label{fig_stability}
\end{center}
\end{figure}

The analysis of FORS2 linear polarization stability and accuracy is based on eight unpolarized standards stars, observed at 40 epochs in total, between 2010 and 2015 (see Table~\ref{tab:stars}). For all stars and each epoch, we calculated the mean value of the Stokes parameters $P_Q$ and $P_U$ in the  3950-9300 \AA\/ wavelength range (Passband "all" in Table~\ref{tbunpol}), which is shown in Figure~\ref{fig_stability}. The values do not show statistically significant variations over the full time range, with very similar standard deviations of $P_Q$ and $P_U$. However, a small but statistically significant instrumental offset is detected in $P_Q$. The weighted mean of the Stokes parameter $P_Q$ of all stars and epochs is 0.07$\pm$0.01\%, while the corresponding value for $P_U$ is 0.00$\pm$0.01\%.  

\begin{figure}
\begin{center}
\includegraphics[trim=20mm 40mm 20mm 40mm, width=9cm, clip=true]{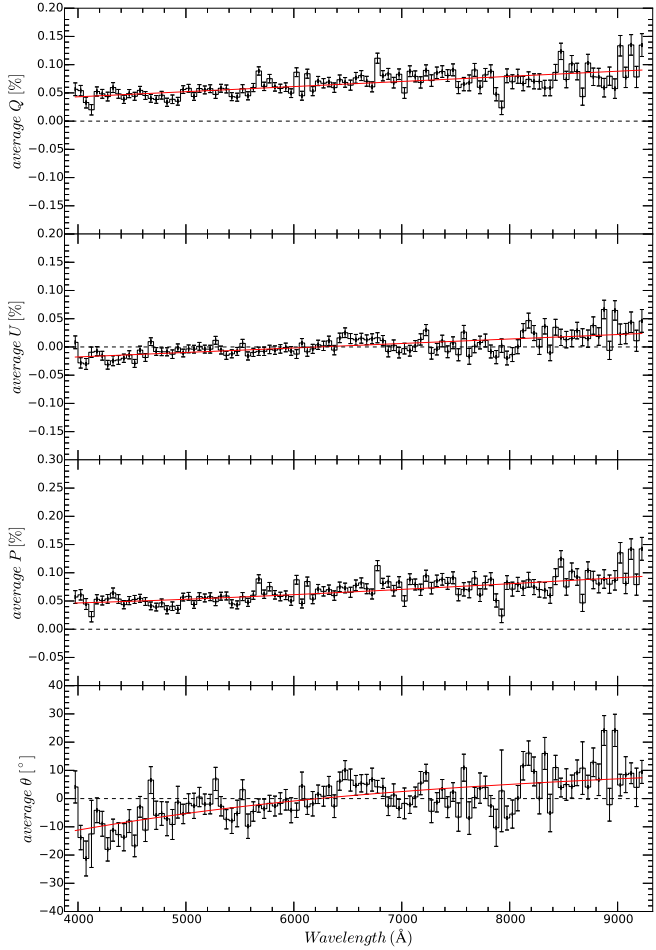}
\vspace{-4mm}
\caption{Weighted means of Stokes parameters Q and U of all observations of unpolarized stars (8 objects, 40 epochs). The red lines are linear least-squares fits to the weighted averages Q and U. The polarization $P$ and polarization angle $\theta$ were calculated from the weighted means of Q and U. The red curves over-plotted on $P$ and $\theta$ were calculated from separate linear fits to $Q(\lambda)$ and $U(\lambda)$.}
\label{fig_allstarsQUav}
\end{center}
\end{figure}

To investigate possible wavelength dependent effects, we calculated the weighted mean of $P_Q$ and $P_U$ as a function of wavelength using all observations of unpolarized standard stars. A clear, wavelength dependent offset is detected in $Q$ (Figure~\ref{fig_allstarsQUav}). The first-order, best fit laws (red lines in Figure~\ref{fig_allstarsQUav}) are as follows: \\
\begin{equation}
\label{eqparametrization}
\begin{array}{l}
Q(\lambda) = [(9.66 \pm 1.04) \times 10^{-8}] \lambda + (3.29 \pm 6.34) \times 10^{-5} \\
 \\
U(\lambda) = [(7.28 \pm 0.90)\times 10^{-8}] \lambda - (4.54 \pm 0.55) \times 10^{-4}
\end{array}
\end{equation}
where $\lambda$ is expressed in \AA. These relations provide a handy correction to be applied to the observed Q and U values. Rigorously speaking, this implicitly assumes that the correction is additive and does not depend on the incoming signal. This approximation probably holds only to first order. The exact prescription can only be obtained with a proper characterization of the Mueller matrices that describe the optical system, which is beyond the scope of this paper.

\citet{2007ASPC..364..503F} performed a similar analysis using observations of standard stars obtained with FORS1 from 1999 to 2005 in both spectropolarimetric (PMOS) and imaging polarimetry (IPOL) mode. For PMOS, they found that the weighted average of all $P_U$ values in PMOS mode are consistent with a null value, but detected an offset of 0.07$\pm$0.01\%, and 0.09$\pm$0.01 \% in $P_Q(B)$ and $P_Q(V)$ respectively. They did not detect any offset in IPOL mode, and concluded that the $P_Q$ offset may be associated with some, but not all, grism and filter combinations.

Our results, based on observations that were taken through the 300V grism (ID: +10) and no order-sorting filters, confirm the findings of \citet{2007ASPC..364..503F} and lead us to conclude that the source of the instrumental polarization causing the offset in $P_Q$ probably resides in the inclined surface of the prism that is coupled to the transmission grating in the FORS1 and FORS2 grisms.

\subsection{Polarized stars}
\label{Polarizedstars}
\begin{figure}
\begin{center}
\includegraphics[trim=0mm 20mm 0mm 20mm, width=9cm, clip=true]{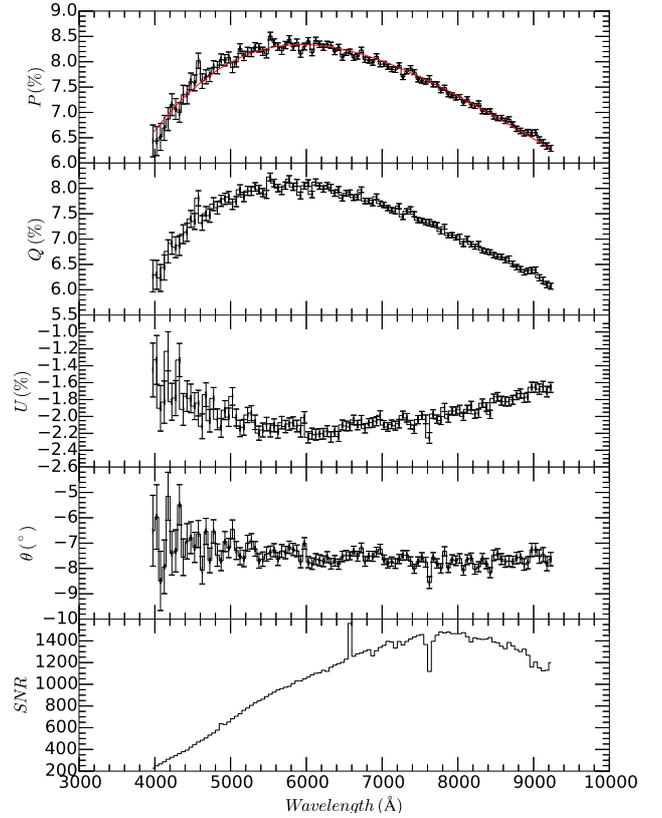}
\vspace{-4mm}
\caption{Total polarization $P$, Stokes parameters $Q$ and $U$, Polarization angle $\theta$ and SNR for Vela1 95 at epoch 2014-01-06. The bin width is 50 \AA. The solid red line is the best fit Serkowski law.}
\label{fig_vela}
\end{center}
\end{figure}

\begin{figure}
\begin{center}
\includegraphics[trim=10mm 10mm 20mm 20mm, width=9cm, clip=true]{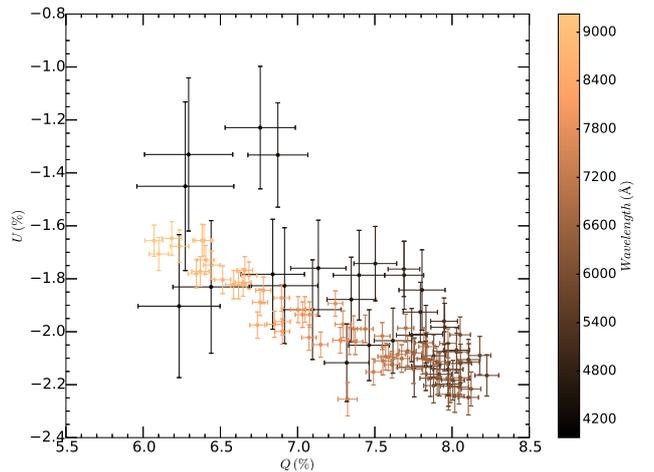}
\vspace{-4mm}
\caption{Q/U diagram for for Vela1 95 at epoch 2014-01-06. The wavelength is color coded.}
\label{fig_velaqu}
\end{center}
\end{figure}

For our analysis we reduced a sample of six highly polarized standard stars (Table~\ref{tab1}) observed with FORS2 between 2009 and 2015 on 30 different epochs. The data were corrected for instrumental polarization using the relations derived from the unpolarised stars (Eqs.~\ref{eqparametrization}). The results for the individual epochs are given in Table~\ref{tbpol}, while weighted means of all epochs are listed in Table~\ref{tb_pol}.
An example of polarization wavelength dependence is shown in Figure~\ref{fig_vela}, and a Q/U diagram in Figure~\ref{fig_velaqu}. As expected for polarization generated by interstellar dust sharing the same alignment angle, the polarization angle is constant, and the points on the Q/U diagram follow a straight line.

{\tiny
\begin{table*}
\centering
\caption{Weighted averages of polarized standard stars observed with FORS2.}
\label{tb_pol}
\begin{tabular}{lclllllll}
\hline\hline
&&&&&& \multicolumn{3}{c}{Serkowski law} \\\cline{7-9}
Name & Passband & $P \hspace{0.3mm} (\%)$ &  $P_Q \hspace{0.3mm} (\%)$ &  $P_U \hspace{0.3mm} (\%)$ & $\theta \hspace{0.3mm} (^{\circ})$ & $\lambda_{\mathrm{max}} \hspace{0.3mm} (\AA)$ &  $P_{\mathrm{max}} \hspace{0.3mm} (\%)$ & $K$ \\
\hline
Vela1 95 &  & &  &   & &   5864.5 $\pm$ 7.4 & 8.295 $\pm$ 0.004 & 1.34 $\pm$ 0.01\\
                    &   \textit{B} &7.645 $\pm$ 0.044 & 7.445 $\pm$ 0.044 & -1.739 $\pm$ 0.046 & 172.76 $\pm$ 0.05& & &\\
                    &   \textit{V} &8.163 $\pm$ 0.011 & 7.834 $\pm$ 0.011 & -2.295 $\pm$ 0.012 & 172.41 $\pm$ 0.02& & &\\
                    &   \textit{R} &7.927 $\pm$ 0.003 & 7.56 $\pm$ 0.003 & -2.383 $\pm$ 0.003 & 172.06 $\pm$ 0.01& & &\\
                    &   \textit{I} &7.151 $\pm$ 0.002 & 6.9 $\pm$ 0.002 & -1.881 $\pm$ 0.002 & 171.95 $\pm$ 0.01& & &\\
BD -14 4922 &  & &  &   & &   5452.5 $\pm$ 13.9 & 6.137 $\pm$ 0.007 & 1.3 $\pm$ 0.02\\
                    &   \textit{B} &5.801 $\pm$ 0.024 & -0.992 $\pm$ 0.024 & 5.715 $\pm$ 0.024 & 49.7 $\pm$ 0.07& & &\\
                    &   \textit{V} &6.096 $\pm$ 0.012 & -0.955 $\pm$ 0.012 & 6.021 $\pm$ 0.012 & 49.8 $\pm$ 0.05& & &\\
                    &   \textit{R} &5.818 $\pm$ 0.006 & -0.864 $\pm$ 0.005 & 5.753 $\pm$ 0.006 & 49.7 $\pm$ 0.04& & &\\
                    &   \textit{I} &4.99 $\pm$ 0.006 & -0.796 $\pm$ 0.006 & 4.926 $\pm$ 0.006 & 49.24 $\pm$ 0.05& & &\\
HDE 316232 &  & &  &   & &   5591.1 $\pm$ 18.3 & 5.017 $\pm$ 0.008 & 1.2 $\pm$ 0.03\\
                    &   \textit{B} &4.679 $\pm$ 0.023 & 4.637 $\pm$ 0.023 & 0.619 $\pm$ 0.021 & 3.61 $\pm$ 0.09& & &\\
                    &   \textit{V} &4.931 $\pm$ 0.014 & 4.901 $\pm$ 0.014 & 0.547 $\pm$ 0.012 & 3.51 $\pm$ 0.07& & &\\
                    &   \textit{R} &4.772 $\pm$ 0.007 & 4.745 $\pm$ 0.007 & 0.508 $\pm$ 0.006 & 3.37 $\pm$ 0.06& & &\\
                    &   \textit{I} &4.214 $\pm$ 0.008 & 4.176 $\pm$ 0.008 & 0.562 $\pm$ 0.007 & 3.53 $\pm$ 0.08& & &\\
Hiltner 652 &  & &  &   & &   5776.5 $\pm$ 9.0 & 6.467 $\pm$ 0.005 & 1.17 $\pm$ 0.01\\
                    &   \textit{B} &5.948 $\pm$ 0.017 & 5.948 $\pm$ 0.017 & -0.054 $\pm$ 0.017 & 179.52 $\pm$ 0.05& & &\\
                    &   \textit{V} &6.371 $\pm$ 0.009 & 6.367 $\pm$ 0.009 & -0.198 $\pm$ 0.009 & 179.44 $\pm$ 0.03& & &\\
                    &   \textit{R} &6.218 $\pm$ 0.004 & 6.214 $\pm$ 0.004 & -0.218 $\pm$ 0.004 & 179.39 $\pm$ 0.03& & &\\
                    &   \textit{I} &5.613 $\pm$ 0.004 & 5.612 $\pm$ 0.004 & -0.041 $\pm$ 0.004 & 179.46 $\pm$ 0.03& & &\\
NGC 2024 1 &  & &  &   & &   6340.4 $\pm$ 4.7 & 9.855 $\pm$ 0.004 & 1.29 $\pm$ 0.01\\
                    &   \textit{B} &8.602 $\pm$ 0.044 & 0.583 $\pm$ 0.045 & -8.582 $\pm$ 0.044 & 136.43 $\pm$ 0.05& & &\\
                    &   \textit{V} &9.548 $\pm$ 0.013 & 0.122 $\pm$ 0.013 & -9.546 $\pm$ 0.013 & 135.94 $\pm$ 0.02& & &\\
                    &   \textit{R} &9.671 $\pm$ 0.004 & -0.0 $\pm$ 0.004 & -9.67 $\pm$ 0.004 & 135.93 $\pm$ 0.01& & &\\
                    &   \textit{I} &9.009 $\pm$ 0.002 & 0.398 $\pm$ 0.002 & -8.999 $\pm$ 0.002 & 135.9 $\pm$ 0.01& & &\\
BD -12 5133 &  & &  &   & &   5049.5 $\pm$ 35.5 & 4.369 $\pm$ 0.01 & 1.2 $\pm$ 0.04\\
                    &   \textit{B} &4.217 $\pm$ 0.027 & 1.688 $\pm$ 0.027 & -3.865 $\pm$ 0.027 & 146.54 $\pm$ 0.12& & &\\
                    &   \textit{V} &4.266 $\pm$ 0.016 & 1.568 $\pm$ 0.016 & -3.968 $\pm$ 0.016 & 145.88 $\pm$ 0.09& & &\\
                    &   \textit{R} &3.996 $\pm$ 0.008 & 1.406 $\pm$ 0.008 & -3.74 $\pm$ 0.008 & 145.62 $\pm$ 0.08& & &\\
                    &   \textit{I} &3.348 $\pm$ 0.009 & 1.222 $\pm$ 0.009 & -3.117 $\pm$ 0.009 & 145.28 $\pm$ 0.11& & &\\
\hline
\end{tabular}
\end{table*}
}

\begin{figure}
\begin{center}
\includegraphics[trim=15mm 5mm 20mm 15mm, width=9cm, clip=true]{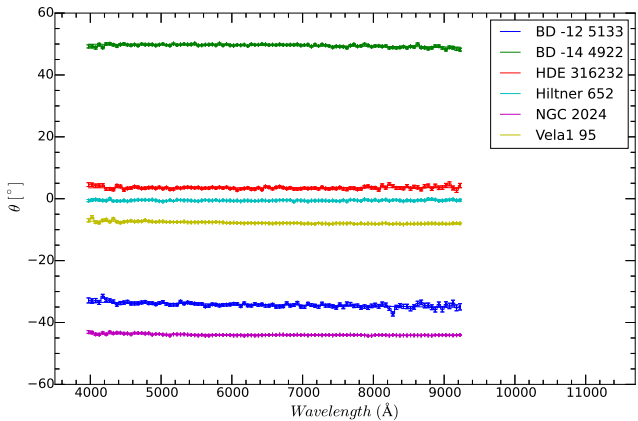}
\vspace{-4mm}
\caption{Weighted mean of polarization angle $\theta$ as a function of wavelength.}
\label{fig_thetalambda}
\end{center}
\end{figure}

The wavelength dependence of the polarization angle was characterized by fitting a second order polynomial to the calculated weighted mean of the polarization angle (Figure~\ref{fig_thetalambda}). From the values of the best fit coefficients we conclude that there is no significant $\theta$ - $\lambda$ dependence. The polarization angles are all constant to within the measurement errors, with slopes d$\theta$/d$\lambda$ between -2.5 and 0.53 degree/$\mu$m.

\begin{figure}
\begin{center}
\includegraphics[trim=20mm 5mm 20mm 20mm, width=9cm, clip=true]{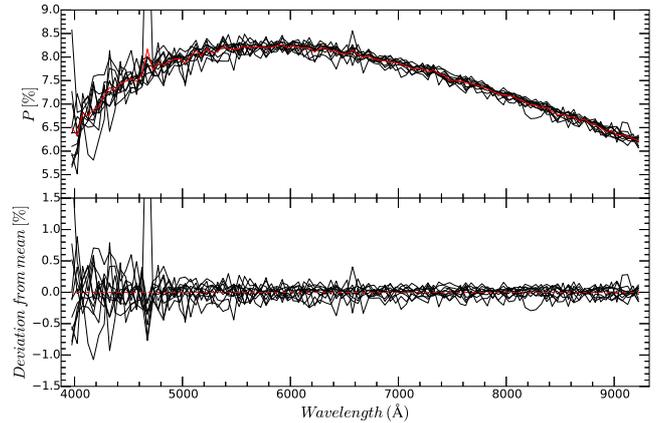}
\vspace{-4mm}
\caption{Reproducibility of Vela1 95 polarization $P$. The bin width is 50 \AA. The solid red line is the mean polarization of all 11 epochs. The deviations form the mean are shown in the bottom plot. The RMS of the deviation from the mean value is 0.21\%.}
\label{fig_vela_rep}
\end{center}
\end{figure}

We tested the reproducibility of the observations using stars observed at least at two epochs: Vela1 95, Hiltner 652, NGC 2024 1 and BD-144922. An example for Vela1 95 is presented in Fig.~\ref{fig_vela_rep}. The root mean square deviation from the mean polarization values are 0.21$\%$ (Vela1 95, 11 epochs), 0.12$\%$ (Hiltner 652, 8 epochs), 0.12\%  (NGC 2024 1, 7 epochs), and  0.05$\%$ (BD-144922, 2 epochs). 

\begin{figure}
\begin{center}
\includegraphics[trim=15mm 0mm 20mm 10mm, width=9cm, clip=true]{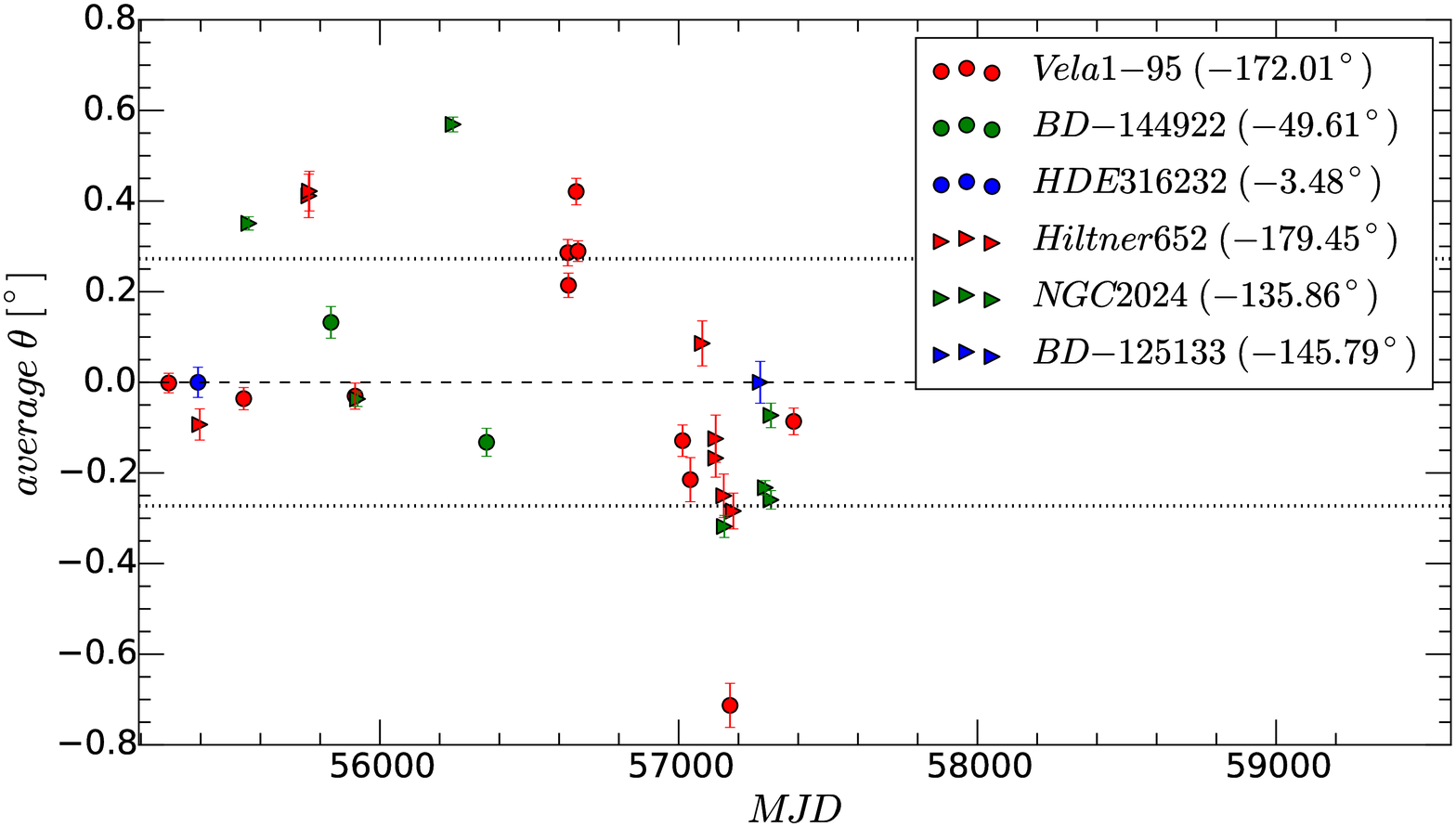}
\vspace{-4mm}
\caption{Polarization angle stability. Shown are residuals of the averaged $\theta(\lambda)$ values for 30 epochs of 6 polarized standard stars.}
\label{fig_stability_theta}
\end{center}
\end{figure}

We finally tested the HWP positioning stability, by calculating the residuals of the mean $\theta(\lambda)$  values for all epochs of the polarized stars.  No systematic jumps are detected, while erratic fluctuations are present, with a peak-to-peak amplitude of about 1.2$^\circ$ and a standard deviation of 0.27$^\circ$ (Figure~\ref{fig_stability_theta}). Given the typical statistical errors of the single measurements, the observed fluctuations are statistically very significant and certainly not due to the photon noise. The most likely interpretation is a possible drift in the absolute positioning of the retarder plate and/or of the analyzer. The estimated rms deviation ($\sim$0.3 degrees) hence represents the typical maximum accuracy one can expect on the polarization angle for very high signal-to-noise ratios.

\subsubsection{Comparison with the literature}

\begin{figure*}
\begin{center}
\includegraphics[trim=35mm 5mm 35mm 25mm, width=18cm, clip=true]{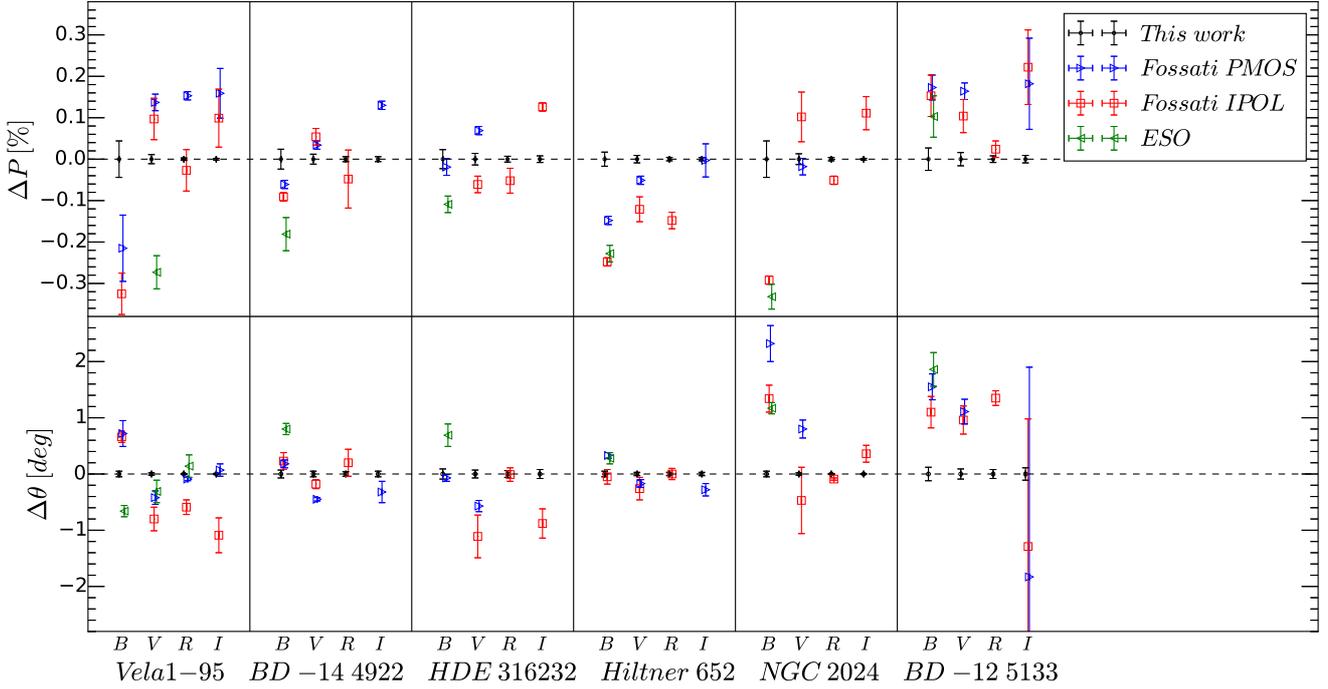}
\vspace{-4mm}
\caption{Comparison between IPOL and PMOS values presented in \citet{2007ASPC..364..503F} (their Table~2), the ESO values given in the FORS webpage, and the new values obtained in this work. $\Delta P$ and $\Delta \theta$ are offsets from values obtained in this work. Three data points are outside of axis limits: the offset, $\Delta P$, from the ESO value for Vela1 95, is 0.60$\pm$0.03$\%$ and -0.76$\pm$0.04$\%$ in \textit{B} and \textit{R} band respectively, while $\Delta P$ in \textit{B} band, from PMOS value \citep{2007ASPC..364..503F} for NGC 2024 1 is $\Delta P$=-0.51$\pm$0.06$\%$.}
\label{fig_literature_comparision}
\end{center}
\end{figure*}

We compared our synthetic broad-band results (Table~\ref{tb_pol}) to those reported by  \citet{2007ASPC..364..503F} (their Table~2) and the values based on data taken with FORS1 during commissioning, given in the FORS webpage\footnote{$http://www.eso.org/sci/facilities/paranal/instruments/fors/inst/\\pola.html$}. Figure~\ref{fig_literature_comparision} shows the polarization and polarization angle differences, $\Delta P$ and $\Delta \theta$, between our measurements and those published in the literature:
$\Delta P$ = $P_{\mathrm{Literature}}$ - $P_{\mathrm{This \hspace{1 pt} work}}$ and $\Delta \theta$ = $\theta_{\mathrm{Literature}}$ - $\theta_{\mathrm{This \hspace{1 pt} work}}$.

The average polarization deviation of all values, $\langle \Delta P \rangle$, obtained in PMOS mode by \citet{2007ASPC..364..503F} from values obtained in this work is 0.01$\%$, with a root mean square deviation of 0.17$\%$. Also the values obtained by \citet{2007ASPC..364..503F} in IPOL mode are consistent with our results, with an average $\langle \Delta P \rangle$ of -0.02$\%$ and an RMS of 0.15$\%$. Our results are on average larger than the values given in the FORS webpage, with $\langle \Delta P \rangle$ = -0.15$\%$ and an RMS of 0.36$\%$.

For the polarization angle, the average deviation from the PMOS values given in \citet{2007ASPC..364..503F} is $\langle \Delta \theta \rangle$=0.17$^\circ$, with an RMS of 0.91$^\circ$; $\langle \Delta\theta \rangle$=-0.03$^\circ$, with an RMS of 0.77$^\circ$ for IPOL mode values \citep{2007ASPC..364..503F}; and $\langle \Delta \theta \rangle$=0.50$^\circ$, with an RMS of 0.76$^\circ$ for the values given in the FORS webpage.

Thus, our results are, in general, in good agreement with previous estimates. The discrepancies with results given in \citet{2007ASPC..364..503F} are consistent with the RMS variation found in the repeatability test (RMS $\lesssim$ 0.2$\%$, see $\S$~\ref{Polarizedstars}), and might be due to star variability or instrumental effects (e.g. the HWP positioning uncertainty, see Figure~\ref{fig_stability_theta}), while the discrepancy from the values given in the FORS webpage is probably due to systematic differences between FORS2 and FORS1.

%
%

\begin{figure}
\begin{center}
\includegraphics[trim=15mm 5mm 5mm 15mm, width=9cm, clip=true]{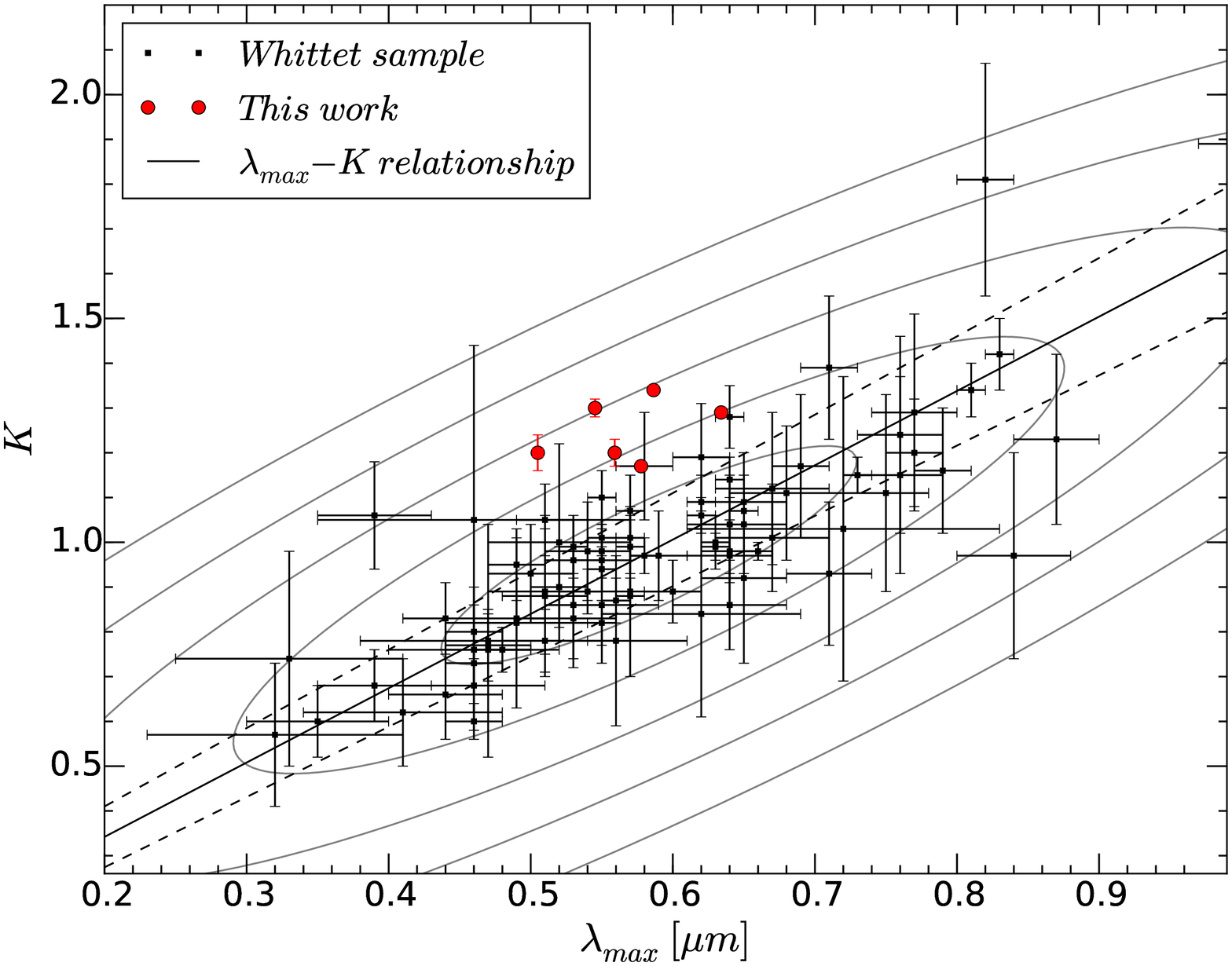}
\vspace{-4mm}
\caption{$\lambda_{\mathrm{max}} - K$ plane containing our 6 standard stars (red dots) compared to the sample from \citet{1992ApJ...386..562W} (black dots). The full line is the empirical $\lambda_{\mathrm{max}} - K$ relationship determined in \citet{1992ApJ...386..562W}, and the dashed lines trace the 1$\sigma$ uncertainty. The gray curves are 1 to 5 sigma error ellipses.}
\label{fig_Whittet_mine}
\end{center}
\end{figure}

We also compared our Serkowski law fitting results for the 6 polarized standard stars to a data set of 105 objects studied by \citet{1992ApJ...386..562W}. They collected data obtained polarimetric observation in \textit{UBVRIJHK} passbands using the Hartfield polarimeter (described by \citet{1986MNRAS.221..739B}, and \citet{1982PASP...94..618B}) on the 3.9-m AAT at Siding Spring Observatory, and the 3.8-m UKIRT at Mauna Kea Observatory. As shown in Figure~\ref{fig_Whittet_mine}, our results deviate by 2-3$\sigma$ from the best fit $\lambda_{\mathrm{max}} - K$ relation found by \citet{1992ApJ...386..562W}.

This deviation, which is statistically significant, cannot be explained by measurement errors, and calls for further investigation. For this reasons, we have looked in more detail the case of Vela1~95, which  is included in the Whittet et al. sample. For this object, \citet{1992ApJ...386..562W} determined a $\lambda_{\mathrm{max}}$=5500$\pm$200 \AA, $P_{\mathrm{max}}$=8.08$\pm$0.07 \% and $K$=1.10$\pm$0.06. From our data, we determined $\lambda_{\mathrm{max}}$=5864$\pm$7 \AA, $P_{\mathrm{max}}$=8.295$\pm$0.004 \% and $K$=1.34$\pm$0.01. 

\begin{figure}
\begin{center}
\includegraphics[trim=25mm 5mm 5mm 15mm, width=9cm, clip=true]{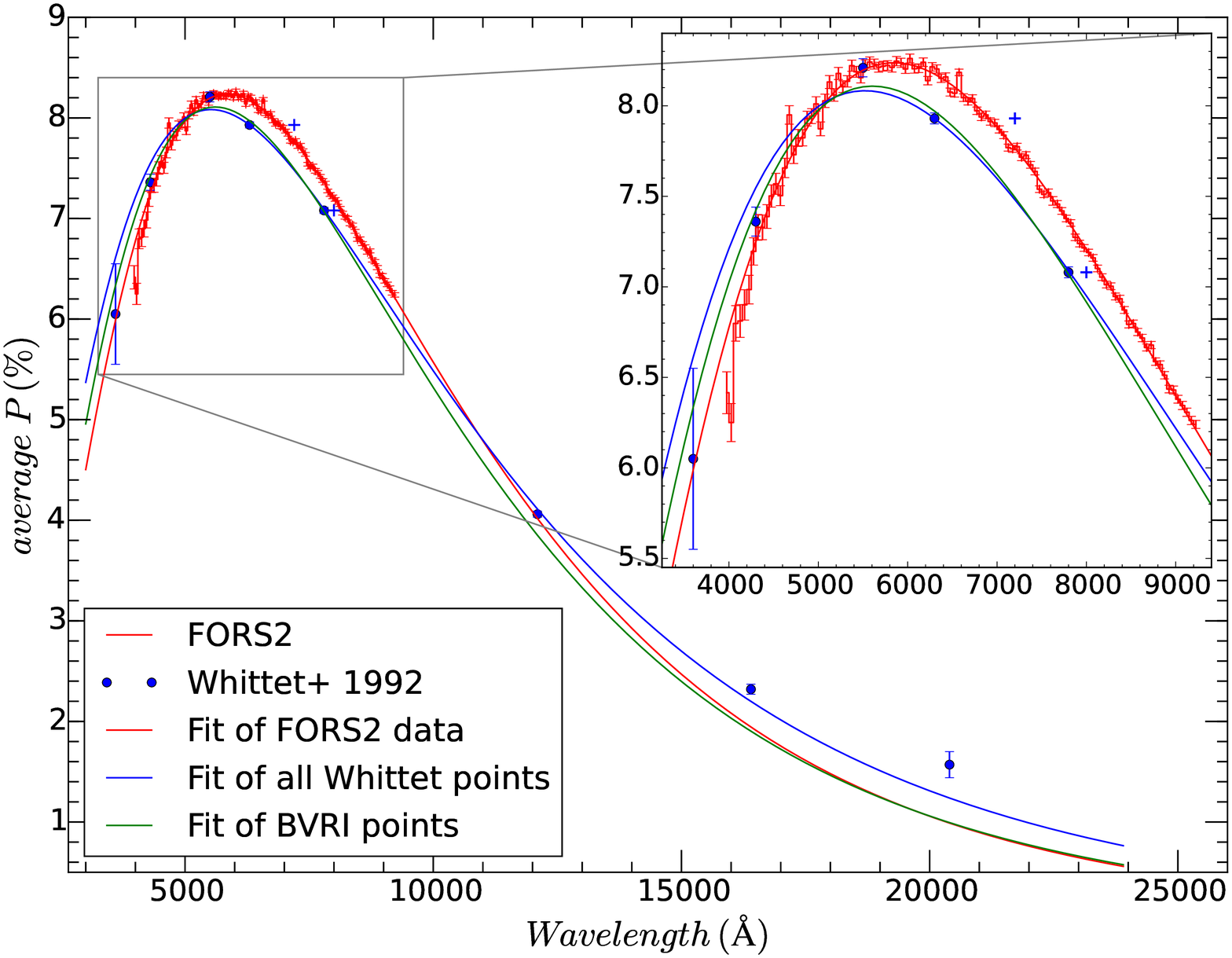}
\vspace{-4mm}
\caption{Comparison between \citet{1992ApJ...386..562W} polarimetric measurements  (blue dots) and spectropolarimetric FORS2 data (red line) for Vela1 95. The colored curves trace a best fit Serkowski law to all data points (blue), \textit{BVRI} data points only (green) and FORS2 data (red). Blue '+' signs mark the effective wavelengths of \textit{R} (0.72 $\mu$m) and \textit{I} (0.80 $\mu$m) passbands, as specified in \citet{1986MNRAS.221..739B} and \citet{1982PASP...94..618B}.}
\label{fig_Whittet_comp}
\end{center}
\end{figure}

In Figure~\ref{fig_Whittet_comp} we compare the \citet{1992ApJ...386..562W} measurements of Vela1 95 to our FORS2 measurements. The figure also includes the Serkowski law best fit to Whittet et al.'s data ($\lambda_{\mathrm{max}}$=5521$\pm$111 \AA, $P_{\mathrm{max}}$=8.08$\pm$0.07 \% and $K$=1.10$\pm$0.05; which is consistent to the Serkowski fit parameters in \citet{1992ApJ...386..562W}), compared to a fit to FORS2 data ($\lambda_{\mathrm{max}}$=5864$\pm$7 \AA, $P_{\mathrm{max}}$ = 8.295$\pm$0.004 \% and $K$=1.34$\pm$0.01), and Whittet et al. \textit{BVRI} data points only ($\lambda_{\mathrm{max}}$=5606$\pm$ 126 \AA, $P_{\mathrm{max}}$=8.11 $\pm$ 0.08 \% and $K$=1.25$\pm$0.19). While the FORS2 data perfectly match the \textit{B} and \textit{V} measurements by Whittet et al., a significant difference is seen in \textit{R} and \textit{I} passbands. This strongly suggests that the problem does not reside in the different wavelength ranges used in the two studies.
The explanation for the observed discrepancy is most likely related to the effective wavelengths adopted by \citet{1992ApJ...386..562W} in their work. The authors list the bandpass properties in their Table~1. They also remark that their $K$ measurements are at an effective wavelength of 2.04 $\mu$m rather than at the usual value of 2.2$\mu$m. This is justified by an absorption in the Foster prism, which narrows the passband and reduces the effective wavelength. However, the authors do not explain why they do not adopt the usual \textit{R} and \textit{I} values given in the instrument description (\citet{1986MNRAS.221..739B}; \citet{1982PASP...94..618B}). The central wavelengths reported by \citet{1992ApJ...386..562W} are: 0.36 $\mu$m (\textit{U}), 0.43 $\mu$m (\textit{B}), 0.55 $\mu$m (\textit{V}), 0.63 $\mu$m (\textit{R}), 0.78 $\mu$m (\textit{I}), 1.21 $\mu$m (\textit{J}), 1.64 $\mu$m (\textit{H}) and 2.04 $\mu$m (\textit{K}). On the other hand, \citet{1986MNRAS.221..739B} list the following wavelengths: 0.36 $\mu$m (\textit{U}), 0.43 $\mu$m (\textit{B}), 0.55 $\mu$m (\textit{V}), 0.72 $\mu$m (\textit{R}), 0.80 $\mu$m (\textit{I}), 1.2 $\mu$m (\textit{J}), 1.64 $\mu$m (\textit{H}) and 2.19 $\mu$m (\textit{K}), which are also consistent with wavelengths in \citet{1982PASP...94..618B}, except that the latter specify 2.14 $\mu$m for the effective wavelength of the K passband. The difference in the \textit{R} and \textit{I} effective wavelengths is evident.

When using 0.72 $\mu$m and 0.80 $\mu$m for \textit{R} and \textit{I} respectively, a Serkowski law fit to all polarimetric points gives: $\lambda_{\mathrm{max}}$=5732$\pm$160 \AA, $P_{\mathrm{max}}$=8.28$\pm$0.12 \% and $K$=1.24$\pm$0.09. When fitting the \textit{BVRI} data only, the best fit parameters are: $\lambda_{\mathrm{max}}$=5834$\pm$168 \AA, $P_{\mathrm{max}}$=8.42$\pm$0.22 \% and $K$=1.65$\pm$0.45. The results are summarized in Table~\ref{tab:serkowsisummary}.
Since spectropolarimetric data do not suffer from the additional problem caused by the need of properly characterizing the photometric system, and given the superior quality of the data presented here and its higher signal to noise, we tend to believe our results are very robust and provide a solid reference.
On these grounds, we suspect that the $\lambda_{\mathrm{max}}$, $P_{\mathrm{max}}$ and $K$ values reported by \citet{1992ApJ...386..562W} are systematically smaller than real. We will further investigate this problem and its consequences in a separate study, using observations along the lines of sight to a larger sample of Galactic reddened stars.

{\tiny
\begin{table*}
\centering
\caption{\label{tab:serkowsisummary} Serkowski parameters comparison table.}
\begin{tabular}{llll}
\hline\hline
                       & $\lambda_{\mathrm{max}}$ ($\AA$)	& $P_{\mathrm{max}}$ ($\%$) & $K$	\\
\hline
\citet{1992ApJ...386..562W} & 5500 $\pm$ 200 & 8.08 $\pm$ 0.07   & 1.10 $\pm$ 0.06 \\
Best fit to FORS2 data	& 5864 $\pm$ 7   & 8.295 $\pm$ 0.004 & 1.34 $\pm$ 0.01 \\
Best fit to all Whittet et al. data points & 5521 $\pm$ 111 & 8.08 $\pm$ 0.07 & 1.10 $\pm$ 0.05 \\
Best fit to Whittet et al. \textit{BVRI} data only & 5606 $\pm$ 126 & 8.11 $\pm$ 0.08 & 1.25 $\pm$ 0.19 \\
Best fit to Whittet et al. \textit{UBVR'I'JHK} data &  5732 $\pm$ 160  &  8.28 $\pm$ 0.12 & 1.24 $\pm$ 0.09 \\
Best fit to Whittet et al. \textit{BVR'I'} data only &  5834 $\pm$ 168& 8.42 $\pm$ 0.22& 1.65 $\pm$ 0.45 \\
\hline
\multicolumn{4}{l}{\textit{R'} and \textit{I'} indicate the modified passband wavelengths, 0.72 $\mu$m and 0.80 $\mu$m for \textit{R} and \textit{I} respectively.}\\
\end{tabular}
\end{table*}
}

\section{Summary and conclusions}

In this study we used archival FORS2 observations of polarimetric standard stars to characterize the performance and stability of FORS2 mounted at the VLT. For this purpose we analyzed 8 unpolarized standard stars observed on 40 epochs, and 6 polarized standard stars (30 epochs). Our main results can be summarized as follows:

\begin{enumerate}
\item We confirm the instrumental wavelength dependent polarization detected by \citet{2007ASPC..364..503F}. The spurious signal steadily grows from the blue to the red, ranging from 0.05\% (4000\AA) to 0.10\% (9000\AA). The vectorial correction for $P_Q$ and $P_U$ (see Equation~\ref{eqparametrization}) can be applied to the observed Stokes parameters. The physical cause of this instrumental polarization is still unclear, but it is most likely related to the tilted surfaces of the despersive element.
\item We tested the repeatability of total linear polarization using observations of polarized standard stars spanning 5 years. The RMS variation is $\lesssim$ 0.2$\%$. For comparison, the typical error per 50$\AA$ bin for a single epoch is $\lesssim$ 0.1$\%$.
\item Using the same sample of polarized stars, we tested the HWP positioning stability, and found a RMS of 0.27$^\circ$. For comparison, the typical uncertainty in the weighted mean is $\sigma_{\theta}$ $\lesssim$ 0.05$^{\circ}$, while the typical uncertainty per 50$\AA$ bin for a single epoch is $\lesssim$ 0.5$^{\circ}$. 
\item Our analysis confirms that FORS2 can achieve a maximum accuracy of $\sim$0.1\% and $\sim$0.3 degrees in polarization degree and angle, respectively. These values, which represent the instrumental limits of this focal reducer, are shown to be stable over timescales of years.
\item As a  by-product of our analysis, we studied the wavelength dependence of linear polarization for our set of polarized standard stars. Our polarization results are in good agreement with those reported in the literature (see Figure~\ref{fig_literature_comparision}). Peak polarization ($P_{\mathrm{max}}$), peak wavelength ($\lambda_{\mathrm{max}}$) and width parameter ($K$) were determined via least squares fit of the classical Serkowski law to the observed data.  
\item Given the larger telescope size and the superb performance of FORS2, the data presented in this paper provide a robust re-calibration of a selected set of linear polarization standard stars that can be reliably used for both checking the performance of other polarimeters and calibrating scientific data obtained with those instruments.
\item A comparison between the $\lambda_{\mathrm{max}}$-$K$ values presented in this paper and those reported by \citet{1992ApJ...386..562W} reveals a systematic and statistically significant deviation, with our $K$ values being larger. While this is partially caused by the shorter wavelength range covered by our FORS2 observations (4000-9000\AA\/ to be compared to the \textit{UBVRIJHK} measurements by Whitthet et al.), it is clear that this cannot fully explain the observed differences. A closer investigation shows that the $K$ values reported by \citet{1992ApJ...386..562W} are most likely offset because of incorrect effective wavelengths for \textit{R} and \textit{I} passbands. This is clearly visible when comparing the data for the standard star Vela1~95, which is common to both data sets.
\item We expect that further studies, including larger sets of reddened Galactic stars, will show the same systematic discrepancy.
\end{enumerate}

\section*{Acknowledgements}
Based on observations made with ESO Telescopes at the Paranal Observatory under programme ID 60.A-9800, 60.A-9203, 084.B-0217, 084.D-0799, 085.D-0391, 089.D-0515, 091.D-0401, 290.D-5009.

\bibliographystyle{mnras} 
\bibliography{specpol.bib} 


\onecolumn

{\tiny
\begin{longtable}{llcccl}
\caption{\label{tbunpol} Unpolarized stars.}\\
\hline\hline
Name     & Epoch   		      & Passband &  $P_Q \hspace{0.3mm} (\%)$ &  $P_U \hspace{0.3mm} (\%)$ & S/N  \\
\hline
\endfirsthead
\caption{continued.}\\
\hline\hline
Name     & Epoch   		      & Passband &  $P_Q \hspace{0.3mm} (\%)$ &  $P_U \hspace{0.3mm} (\%)$ & S/N  \\
\hline
\endhead
\hline
\endfoot
HD10038& 2010-09-24  & all & 0.076 $\pm$ 0.05 & -0.025 $\pm$ 0.052& \\ 
         &           &   \textit{B} &0.065 $\pm$ 0.012 & -0.02 $\pm$ 0.013& 1553\\ 
         &           &   \textit{V} &0.065 $\pm$ 0.009 & -0.025 $\pm$ 0.01& 1927\\ 
         &           &   \textit{R} &0.079 $\pm$ 0.005 & -0.034 $\pm$ 0.006& 1726\\ 
         &           &   \textit{I} &0.077 $\pm$ 0.007 & -0.032 $\pm$ 0.008& 1282\\ 
HD10038& 2010-10-02  & all & 0.027 $\pm$ 0.084 & -0.002 $\pm$ 0.08& \\ 
         &           &   \textit{B} &0.023 $\pm$ 0.022 & 0.027 $\pm$ 0.021& 908\\ 
         &           &   \textit{V} &0.012 $\pm$ 0.016 & 0.011 $\pm$ 0.015& 1174\\ 
         &           &   \textit{R} &0.023 $\pm$ 0.009 & 0.008 $\pm$ 0.009& 1062\\ 
         &           &   \textit{I} &0.013 $\pm$ 0.012 & -0.008 $\pm$ 0.011& 799\\ 
HD10038& 2011-12-10  & all & 0.041 $\pm$ 0.061 & -0.048 $\pm$ 0.062& \\ 
         &           &   \textit{B} &0.031 $\pm$ 0.014 & -0.055 $\pm$ 0.014& 1317\\ 
         &           &   \textit{V} &0.035 $\pm$ 0.011 & -0.053 $\pm$ 0.012& 1608\\ 
         &           &   \textit{R} &0.049 $\pm$ 0.007 & -0.064 $\pm$ 0.007& 1428\\ 
         &           &   \textit{I} &0.045 $\pm$ 0.009 & -0.042 $\pm$ 0.009& 1050\\ 
HD10038& 2012-12-25  & all & 0.043 $\pm$ 0.086 & 0.021 $\pm$ 0.084& \\ 
         &           &   \textit{B} &0.045 $\pm$ 0.021 & -0.0 $\pm$ 0.02& 917\\ 
         &           &   \textit{V} &0.046 $\pm$ 0.016 & 0.005 $\pm$ 0.016& 1152\\ 
         &           &   \textit{R} &0.051 $\pm$ 0.009 & 0.023 $\pm$ 0.009& 1032\\ 
         &           &   \textit{I} &0.044 $\pm$ 0.012 & 0.038 $\pm$ 0.012& 768\\ 
HD10038& 2013-12-14  & all & 0.052 $\pm$ 0.061 & -0.031 $\pm$ 0.062& \\ 
         &           &   \textit{B} &0.041 $\pm$ 0.014 & -0.062 $\pm$ 0.014& 1334\\ 
         &           &   \textit{V} &0.046 $\pm$ 0.011 & -0.033 $\pm$ 0.012& 1608\\ 
         &           &   \textit{R} &0.055 $\pm$ 0.007 & -0.032 $\pm$ 0.007& 1424\\ 
         &           &   \textit{I} &0.059 $\pm$ 0.009 & -0.007 $\pm$ 0.009& 1045\\ 
HD10038& 2015-07-31  & all & 0.04 $\pm$ 0.053 & -0.029 $\pm$ 0.054& \\ 
         &           &   \textit{B} &0.037 $\pm$ 0.014 & -0.034 $\pm$ 0.014& 1378\\ 
         &           &   \textit{V} &0.032 $\pm$ 0.011 & -0.023 $\pm$ 0.011& 1760\\ 
         &           &   \textit{R} &0.041 $\pm$ 0.006 & -0.028 $\pm$ 0.006& 1600\\ 
         &           &   \textit{I} &0.032 $\pm$ 0.008 & -0.025 $\pm$ 0.008& 1230\\ 
HD10038& 2015-08-28  & all & 0.078 $\pm$ 0.055 & -0.063 $\pm$ 0.053& \\ 
         &           &   \textit{B} &0.03 $\pm$ 0.013 & -0.07 $\pm$ 0.013& 1459\\ 
         &           &   \textit{V} &0.065 $\pm$ 0.01 & -0.056 $\pm$ 0.01& 1806\\ 
         &           &   \textit{R} &0.083 $\pm$ 0.006 & -0.06 $\pm$ 0.006& 1621\\ 
         &           &   \textit{I} &0.094 $\pm$ 0.008 & -0.046 $\pm$ 0.008& 1214\\ 
HD10038& 2015-10-14  & all & 0.041 $\pm$ 0.068 & -0.117 $\pm$ 0.064& \\ 
         &           &   \textit{B} &-0.003 $\pm$ 0.017 & -0.117 $\pm$ 0.016& 1185\\ 
         &           &   \textit{V} &0.053 $\pm$ 0.013 & -0.09 $\pm$ 0.012& 1478\\ 
         &           &   \textit{R} &0.063 $\pm$ 0.007 & -0.108 $\pm$ 0.007& 1331\\ 
         &           &   \textit{I} &0.054 $\pm$ 0.01 & -0.146 $\pm$ 0.009& 1002\\ 
HD13588& 2014-10-02  & all & 0.11 $\pm$ 0.056 & 0.034 $\pm$ 0.055& \\ 
         &           &   \textit{B} &0.154 $\pm$ 0.014 & 0.001 $\pm$ 0.014& 1360\\ 
         &           &   \textit{V} &0.101 $\pm$ 0.011 & 0.021 $\pm$ 0.011& 1751\\ 
         &           &   \textit{R} &0.106 $\pm$ 0.006 & 0.05 $\pm$ 0.006& 1586\\ 
         &           &   \textit{I} &0.113 $\pm$ 0.008 & 0.048 $\pm$ 0.008& 1196\\ 
HD13588& 2015-08-14  & all & 0.151 $\pm$ 0.075 & -0.011 $\pm$ 0.078& \\ 
         &           &   \textit{B} &0.099 $\pm$ 0.017 & -0.058 $\pm$ 0.017& 1054\\ 
         &           &   \textit{V} &0.147 $\pm$ 0.014 & 0.002 $\pm$ 0.015& 1273\\ 
         &           &   \textit{R} &0.15 $\pm$ 0.008 & -0.02 $\pm$ 0.009& 1127\\ 
         &           &   \textit{I} &0.165 $\pm$ 0.011 & -0.01 $\pm$ 0.011& 831\\ 
HD13588& 2015-08-26  & all & 0.067 $\pm$ 0.068 & 0.047 $\pm$ 0.068& \\ 
         &           &   \textit{B} &0.049 $\pm$ 0.015 & 0.027 $\pm$ 0.015& 1635\\ 
         &           &   \textit{V} &0.057 $\pm$ 0.013 & 0.038 $\pm$ 0.013& 1952\\ 
         &           &   \textit{R} &0.067 $\pm$ 0.007 & 0.042 $\pm$ 0.007& 1713\\ 
         &           &   \textit{I} &0.074 $\pm$ 0.01 & 0.057 $\pm$ 0.01& 1249\\ 
HD13588& 2015-10-14  & all & 0.149 $\pm$ 0.082 & 0.055 $\pm$ 0.087& \\ 
         &           &   \textit{B} &0.084 $\pm$ 0.018 & 0.029 $\pm$ 0.019& 1303\\ 
         &           &   \textit{V} &0.134 $\pm$ 0.016 & 0.04 $\pm$ 0.016& 1573\\ 
         &           &   \textit{R} &0.164 $\pm$ 0.009 & 0.048 $\pm$ 0.01& 1389\\ 
         &           &   \textit{I} &0.183 $\pm$ 0.012 & 0.081 $\pm$ 0.013& 1023\\ 
HD97689& 2011-02-04  & all & 0.063 $\pm$ 0.092 & -0.028 $\pm$ 0.087& \\ 
         &           &   \textit{B} &0.036 $\pm$ 0.021 & -0.051 $\pm$ 0.02& 912\\ 
         &           &   \textit{V} &0.057 $\pm$ 0.017 & -0.018 $\pm$ 0.016& 1123\\ 
         &           &   \textit{R} &0.07 $\pm$ 0.01 & -0.006 $\pm$ 0.009& 996\\ 
         &           &   \textit{I} &0.065 $\pm$ 0.013 & -0.014 $\pm$ 0.013& 733\\ 
HD97689& 2011-02-12  & all & 0.051 $\pm$ 0.057 & 0.017 $\pm$ 0.055& \\ 
         &           &   \textit{B} &0.047 $\pm$ 0.012 & -0.014 $\pm$ 0.012& 1533\\ 
         &           &   \textit{V} &0.038 $\pm$ 0.01 & 0.003 $\pm$ 0.01& 1843\\ 
         &           &   \textit{R} &0.05 $\pm$ 0.006 & 0.007 $\pm$ 0.006& 1613\\ 
         &           &   \textit{I} &0.044 $\pm$ 0.008 & 0.048 $\pm$ 0.008& 1154\\ 
HD97689& 2011-06-30  & all & 0.275 $\pm$ 0.056 & 0.042 $\pm$ 0.056& \\ 
         &           &   \textit{B} &0.183 $\pm$ 0.013 & 0.014 $\pm$ 0.012& 1470\\ 
         &           &   \textit{V} &0.235 $\pm$ 0.01 & 0.029 $\pm$ 0.01& 1781\\ 
         &           &   \textit{R} &0.266 $\pm$ 0.006 & 0.043 $\pm$ 0.006& 1571\\ 
         &           &   \textit{I} &0.333 $\pm$ 0.008 & 0.066 $\pm$ 0.008& 1147\\ 
HD97689& 2012-01-20  & all & 0.159 $\pm$ 0.077 & -0.012 $\pm$ 0.062& \\ 
         &           &   \textit{B} &0.113 $\pm$ 0.019 & 0.01 $\pm$ 0.015& 1152\\ 
         &           &   \textit{V} &0.144 $\pm$ 0.015 & -0.006 $\pm$ 0.012& 1442\\ 
         &           &   \textit{R} &0.138 $\pm$ 0.008 & -0.001 $\pm$ 0.007& 1294\\ 
         &           &   \textit{I} &0.173 $\pm$ 0.011 & -0.039 $\pm$ 0.009& 960\\ 
HD97689& 2012-03-30  & all & 0.104 $\pm$ 0.062 & 0.051 $\pm$ 0.068& \\ 
         &           &   \textit{B} &0.089 $\pm$ 0.014 & 0.019 $\pm$ 0.016& 1232\\ 
         &           &   \textit{V} &0.085 $\pm$ 0.012 & 0.025 $\pm$ 0.013& 1522\\ 
         &           &   \textit{R} &0.104 $\pm$ 0.007 & 0.047 $\pm$ 0.007& 1356\\ 
         &           &   \textit{I} &0.122 $\pm$ 0.009 & 0.084 $\pm$ 0.01& 999\\ 
HD97689& 2013-05-06  & all & 0.062 $\pm$ 0.055 & -0.022 $\pm$ 0.053& \\ 
         &           &   \textit{B} &0.071 $\pm$ 0.013 & -0.008 $\pm$ 0.012& 1482\\ 
         &           &   \textit{V} &0.064 $\pm$ 0.01 & -0.041 $\pm$ 0.01& 1835\\ 
         &           &   \textit{R} &0.065 $\pm$ 0.006 & -0.02 $\pm$ 0.006& 1631\\ 
         &           &   \textit{I} &0.063 $\pm$ 0.008 & -0.039 $\pm$ 0.008& 1201\\ 
HD97689& 2014-01-06  & all & 0.095 $\pm$ 0.059 & 0.026 $\pm$ 0.06& \\ 
         &           &   \textit{B} &0.065 $\pm$ 0.014 & -0.015 $\pm$ 0.014& 1343\\ 
         &           &   \textit{V} &0.075 $\pm$ 0.011 & 0.011 $\pm$ 0.011& 1661\\ 
         &           &   \textit{R} &0.096 $\pm$ 0.006 & 0.026 $\pm$ 0.006& 1483\\ 
         &           &   \textit{I} &0.109 $\pm$ 0.008 & 0.058 $\pm$ 0.009& 1094\\ 
HD97689& 2014-12-24  & all & 0.033 $\pm$ 0.08 & -0.028 $\pm$ 0.081& \\ 
         &           &   \textit{B} &0.008 $\pm$ 0.018 & -0.015 $\pm$ 0.018& 1022\\ 
         &           &   \textit{V} &0.018 $\pm$ 0.015 & -0.018 $\pm$ 0.015& 1229\\ 
         &           &   \textit{R} &0.033 $\pm$ 0.009 & -0.045 $\pm$ 0.009& 1082\\ 
         &           &   \textit{I} &0.036 $\pm$ 0.012 & -0.035 $\pm$ 0.012& 794\\ 
HD97689& 2015-02-15  & all & 0.099 $\pm$ 0.059 & 0.035 $\pm$ 0.06& \\ 
         &           &   \textit{B} &0.055 $\pm$ 0.013 & 0.017 $\pm$ 0.013& 1387\\ 
         &           &   \textit{V} &0.043 $\pm$ 0.011 & 0.025 $\pm$ 0.011& 1705\\ 
         &           &   \textit{R} &0.066 $\pm$ 0.006 & 0.03 $\pm$ 0.006& 1501\\ 
         &           &   \textit{I} &0.143 $\pm$ 0.009 & 0.016 $\pm$ 0.009& 1092\\ 
HD97689& 2015-04-07  & all & -0.01 $\pm$ 0.057 & 0.053 $\pm$ 0.058& \\ 
         &           &   \textit{B} &-0.012 $\pm$ 0.014 & 0.042 $\pm$ 0.014& 1334\\ 
         &           &   \textit{V} &0.001 $\pm$ 0.011 & 0.054 $\pm$ 0.011& 1691\\ 
         &           &   \textit{R} &-0.004 $\pm$ 0.006 & 0.048 $\pm$ 0.006& 1523\\ 
         &           &   \textit{I} &-0.024 $\pm$ 0.008 & 0.048 $\pm$ 0.008& 1141\\ 
HD97689& 2015-04-07  & all & 0.009 $\pm$ 0.084 & -0.078 $\pm$ 0.092& \\ 
         &           &   \textit{B} &-0.041 $\pm$ 0.023 & -0.066 $\pm$ 0.025& 829\\ 
         &           &   \textit{V} &-0.018 $\pm$ 0.017 & -0.018 $\pm$ 0.019& 1073\\ 
         &           &   \textit{R} &-0.006 $\pm$ 0.009 & -0.058 $\pm$ 0.01& 985\\ 
         &           &   \textit{I} &0.053 $\pm$ 0.012 & -0.122 $\pm$ 0.013& 776\\ 
HD97689& 2015-04-07  & all & 0.137 $\pm$ 0.077 & 0.089 $\pm$ 0.074& \\ 
         &           &   \textit{B} &0.142 $\pm$ 0.021 & 0.002 $\pm$ 0.02& 976\\ 
         &           &   \textit{V} &0.103 $\pm$ 0.016 & 0.045 $\pm$ 0.015& 1246\\ 
         &           &   \textit{R} &0.135 $\pm$ 0.009 & 0.082 $\pm$ 0.008& 1140\\ 
         &           &   \textit{I} &0.136 $\pm$ 0.011 & 0.151 $\pm$ 0.01& 894\\ 
HD42078& 2010-04-07  & all & 0.024 $\pm$ 0.046 & 0.035 $\pm$ 0.045& \\ 
         &           &   \textit{B} &0.066 $\pm$ 0.01 & 0.035 $\pm$ 0.01& 1854\\ 
         &           &   \textit{V} &0.041 $\pm$ 0.008 & 0.056 $\pm$ 0.008& 2217\\ 
         &           &   \textit{R} &0.046 $\pm$ 0.005 & 0.053 $\pm$ 0.005& 1949\\ 
         &           &   \textit{I} &-0.009 $\pm$ 0.007 & 0.011 $\pm$ 0.007& 1419\\ 
HD42078& 2010-10-14  & all & 0.161 $\pm$ 0.077 & 0.13 $\pm$ 0.072& \\ 
         &           &   \textit{B} &0.132 $\pm$ 0.019 & 0.077 $\pm$ 0.018& 1096\\ 
         &           &   \textit{V} &0.143 $\pm$ 0.015 & 0.056 $\pm$ 0.014& 1367\\ 
         &           &   \textit{R} &0.162 $\pm$ 0.008 & 0.086 $\pm$ 0.008& 1230\\ 
         &           &   \textit{I} &0.191 $\pm$ 0.011 & 0.186 $\pm$ 0.01& 924\\ 
HD42078& 2010-12-13  & all & 0.053 $\pm$ 0.06 & 0.044 $\pm$ 0.061& \\ 
         &           &   \textit{B} &0.051 $\pm$ 0.013 & 0.034 $\pm$ 0.014& 1379\\ 
         &           &   \textit{V} &0.046 $\pm$ 0.011 & 0.026 $\pm$ 0.011& 1661\\ 
         &           &   \textit{R} &0.055 $\pm$ 0.006 & 0.045 $\pm$ 0.007& 1459\\ 
         &           &   \textit{I} &0.061 $\pm$ 0.009 & 0.055 $\pm$ 0.009& 1065\\ 
HD42078& 2010-12-17  & all & 0.092 $\pm$ 0.069 & -0.07 $\pm$ 0.072& \\ 
         &           &   \textit{B} &0.089 $\pm$ 0.015 & -0.077 $\pm$ 0.016& 1183\\ 
         &           &   \textit{V} &0.078 $\pm$ 0.013 & -0.067 $\pm$ 0.013& 1415\\ 
         &           &   \textit{R} &0.101 $\pm$ 0.008 & -0.08 $\pm$ 0.008& 1246\\ 
         &           &   \textit{I} &0.101 $\pm$ 0.01 & -0.063 $\pm$ 0.011& 911\\ 
HD42078& 2014-03-04  & all & 0.173 $\pm$ 0.066 & 0.025 $\pm$ 0.065& \\ 
         &           &   \textit{B} &0.158 $\pm$ 0.009 & -0.059 $\pm$ 0.009& 1604\\ 
         &           &   \textit{V} &0.17 $\pm$ 0.01 & -0.048 $\pm$ 0.009& 1651\\ 
         &           &   \textit{R} &0.204 $\pm$ 0.006 & -0.013 $\pm$ 0.006& 1348\\ 
         &           &   \textit{I} &0.202 $\pm$ 0.011 & 0.087 $\pm$ 0.01& 830\\ 
WD1615-154& 2010-04-02  & all & 0.022 $\pm$ 0.083 & 0.023 $\pm$ 0.084& \\ 
         &           &   \textit{B} &-0.007 $\pm$ 0.011 & -0.006 $\pm$ 0.011& 1440\\ 
         &           &   \textit{V} &-0.028 $\pm$ 0.013 & -0.003 $\pm$ 0.013& 1388\\ 
         &           &   \textit{R} &0.003 $\pm$ 0.009 & 0.029 $\pm$ 0.009& 1088\\ 
         &           &   \textit{I} &0.038 $\pm$ 0.013 & 0.014 $\pm$ 0.013& 720\\ 
WD1620-391& 2010-05-30  & all & 0.12 $\pm$ 0.168 & -0.136 $\pm$ 0.168& \\ 
         &           &   \textit{B} &0.014 $\pm$ 0.019 & -0.059 $\pm$ 0.018& 794\\ 
         &           &   \textit{V} &0.124 $\pm$ 0.024 & -0.01 $\pm$ 0.023& 751\\ 
         &           &   \textit{R} &0.115 $\pm$ 0.017 & 0.01 $\pm$ 0.017& 547\\ 
         &           &   \textit{I} &0.348 $\pm$ 0.028 & -0.34 $\pm$ 0.028& 339\\ 
WD1620-391& 2010-07-14  & all & 0.062 $\pm$ 0.063 & 0.055 $\pm$ 0.065& \\ 
         &           &   \textit{B} &0.042 $\pm$ 0.01 & 0.01 $\pm$ 0.01& 1743\\ 
         &           &   \textit{V} &0.038 $\pm$ 0.01 & 0.028 $\pm$ 0.011& 1818\\ 
         &           &   \textit{R} &0.06 $\pm$ 0.007 & 0.048 $\pm$ 0.007& 1466\\ 
         &           &   \textit{I} &0.078 $\pm$ 0.01 & 0.089 $\pm$ 0.01& 992\\ 
WD1620-391& 2011-02-11  & all & 0.056 $\pm$ 0.068 & -0.011 $\pm$ 0.065& \\ 
         &           &   \textit{B} &0.05 $\pm$ 0.01 & -0.002 $\pm$ 0.01& 1596\\ 
         &           &   \textit{V} &0.053 $\pm$ 0.011 & 0.017 $\pm$ 0.01& 1699\\ 
         &           &   \textit{R} &0.068 $\pm$ 0.007 & 0.004 $\pm$ 0.007& 1372\\ 
         &           &   \textit{I} &0.057 $\pm$ 0.01 & -0.049 $\pm$ 0.01& 922\\ 
WD1620-391& 2014-03-22  & all & 0.016 $\pm$ 0.062 & 0.026 $\pm$ 0.066& \\ 
         &           &   \textit{B} &0.017 $\pm$ 0.009 & 0.004 $\pm$ 0.01& 1746\\ 
         &           &   \textit{V} &0.036 $\pm$ 0.01 & -0.004 $\pm$ 0.011& 1814\\ 
         &           &   \textit{R} &0.047 $\pm$ 0.006 & -0.01 $\pm$ 0.007& 1447\\ 
         &           &   \textit{I} &-0.023 $\pm$ 0.01 & 0.045 $\pm$ 0.01& 968\\ 
WD1620-391& 2014-03-22  & all & 0.009 $\pm$ 0.077 & -0.004 $\pm$ 0.079& \\ 
         &           &   \textit{B} &0.008 $\pm$ 0.011 & -0.011 $\pm$ 0.012& 1396\\ 
         &           &   \textit{V} &-0.004 $\pm$ 0.012 & 0.015 $\pm$ 0.013& 1452\\ 
         &           &   \textit{R} &0.005 $\pm$ 0.008 & 0.016 $\pm$ 0.008& 1160\\ 
         &           &   \textit{I} &0.005 $\pm$ 0.012 & -0.04 $\pm$ 0.012& 778\\ 
WD1620-391& 2014-03-22  & all & -0.036 $\pm$ 0.078 & -0.029 $\pm$ 0.078& \\ 
         &           &   \textit{B} &-0.029 $\pm$ 0.011 & -0.008 $\pm$ 0.011& 1399\\ 
         &           &   \textit{V} &-0.033 $\pm$ 0.013 & -0.029 $\pm$ 0.013& 1455\\ 
         &           &   \textit{R} &-0.049 $\pm$ 0.008 & -0.024 $\pm$ 0.008& 1162\\ 
         &           &   \textit{I} &-0.076 $\pm$ 0.012 & -0.054 $\pm$ 0.012& 778\\ 
WD1620-391& 2015-02-26  & all & 0.018 $\pm$ 0.064 & 0.012 $\pm$ 0.065& \\ 
         &           &   \textit{B} &-0.007 $\pm$ 0.009 & -0.017 $\pm$ 0.009& 1714\\ 
         &           &   \textit{V} &0.001 $\pm$ 0.01 & -0.005 $\pm$ 0.01& 1775\\ 
         &           &   \textit{R} &0.006 $\pm$ 0.007 & -0.009 $\pm$ 0.007& 1415\\ 
         &           &   \textit{I} &0.038 $\pm$ 0.01 & 0.029 $\pm$ 0.01& 942\\ 
WD2039-202& 2013-04-12  & all & 0.079 $\pm$ 0.073 & 0.038 $\pm$ 0.076& \\ 
         &           &   \textit{B} &0.046 $\pm$ 0.012 & -0.068 $\pm$ 0.013& 1308\\ 
         &           &   \textit{V} &0.072 $\pm$ 0.012 & -0.009 $\pm$ 0.013& 1464\\ 
         &           &   \textit{R} &0.073 $\pm$ 0.008 & 0.014 $\pm$ 0.008& 1199\\ 
         &           &   \textit{I} &0.08 $\pm$ 0.011 & 0.05 $\pm$ 0.012& 815\\ 
WD2039-202& 2015-09-08  & all & -0.011 $\pm$ 0.066 & -0.039 $\pm$ 0.066& \\ 
         &           &   \textit{B} &-0.003 $\pm$ 0.01 & -0.038 $\pm$ 0.01& 2143\\ 
         &           &   \textit{V} &-0.008 $\pm$ 0.011 & -0.056 $\pm$ 0.011& 2275\\ 
         &           &   \textit{R} &-0.006 $\pm$ 0.007 & -0.041 $\pm$ 0.007& 1831\\ 
         &           &   \textit{I} &-0.02 $\pm$ 0.01 & -0.039 $\pm$ 0.01& 1234\\ 
WD2149+021& 2011-10-11  & all & -0.048 $\pm$ 0.056 & -0.038 $\pm$ 0.056& \\ 
         &           &   \textit{B} &-0.087 $\pm$ 0.009 & -0.018 $\pm$ 0.009& 1776\\ 
         &           &   \textit{V} &-0.026 $\pm$ 0.009 & -0.057 $\pm$ 0.009& 2011\\ 
         &           &   \textit{R} &-0.029 $\pm$ 0.006 & -0.041 $\pm$ 0.006& 1625\\ 
         &           &   \textit{I} &-0.055 $\pm$ 0.009 & -0.058 $\pm$ 0.009& 1091\\ 
\end{longtable}       
The signal-to-noise ratio, S/N, in individual bands is the average SNR between two 50$\AA$ bins, from 4250-4350$\AA$ (\textit{B} band), 5450-5550$\AA$ (\textit{V}), 6250-6350$\AA$ (\textit{R}) and 7650-7850$\AA$ (\textit{I}).\\
}

{\tiny
\begin{longtable}{lccllllllll}
\caption{\label{tbpol} Polarized stars.}\\
\hline\hline
&&&&&&&& \multicolumn{3}{c}{Serkowski law} \\\cline{9-11}
Name & Epoch & Passband & $P \hspace{0.3mm} (\%)$ &  $P_Q \hspace{0.3mm} (\%)$ &  $P_U \hspace{0.3mm} (\%)$ & $\theta \hspace{0.3mm} (^{\circ})$ & S/N & $\lambda_{\mathrm{max}} \hspace{0.3mm} (\AA)$ &  $P_{\mathrm{max}} \hspace{0.3mm} (\%)$ & $K$ \\\hline
\endfirsthead
\caption{continued.}\\
\hline\hline
&&&&&&&& \multicolumn{3}{c}{Serkowski law} \\\cline{9-11}
Name & Epoch & Passband & $P \hspace{0.3mm} (\%)$ &  $P_Q \hspace{0.3mm} (\%)$ &  $P_U \hspace{0.3mm} (\%)$ & $\theta \hspace{0.3mm} (^{\circ})$ & S/N & $\lambda_{\mathrm{max}} \hspace{0.3mm} (\AA)$ &  $P_{\mathrm{max}} \hspace{0.3mm} (\%)$ & $K$ \\\hline
\endhead
\hline
\endfoot
Vela1-95& 2010-04-07   &    &    &    &       &                  & &  5848.2 $\pm$ 21.9         & 8.36 $\pm$ 0.01 & 1.34 $\pm$ 0.03\\ 
         &                     &   \textit{B} &7.69 $\pm$ 0.1 & 7.48 $\pm$ 0.1 & -1.79 $\pm$ 0.1 & 172.72 $\pm$ 0.13&405& & &\\ 
         &                     &   \textit{V} &8.23 $\pm$ 0.03 & 7.9 $\pm$ 0.03 & -2.3 $\pm$ 0.03 & 172.43 $\pm$ 0.06&965& & &\\ 
         &                     &   \textit{R} &7.99 $\pm$ 0.01 & 7.61 $\pm$ 0.01 & -2.42 $\pm$ 0.01 & 171.97 $\pm$ 0.04&1210& & &\\ 
         &                     &   \textit{I} &7.2 $\pm$ 0.01 & 6.95 $\pm$ 0.01 & -1.91 $\pm$ 0.01 & 171.87 $\pm$ 0.03&1516& & &\\ 
Vela1-95& 2010-12-14   &    &    &    &       &                  & &  5877.1 $\pm$ 22.8         & 8.27 $\pm$ 0.01 & 1.34 $\pm$ 0.03\\ 
         &                     &   \textit{B} &7.58 $\pm$ 0.12 & 7.39 $\pm$ 0.12 & -1.68 $\pm$ 0.14 & 172.7 $\pm$ 0.15&328& & &\\
         &                     &   \textit{V} &8.14 $\pm$ 0.03 & 7.8 $\pm$ 0.03 & -2.32 $\pm$ 0.04 & 172.27 $\pm$ 0.07&801& & &\\ 
         &                     &   \textit{R} &7.91 $\pm$ 0.01 & 7.54 $\pm$ 0.01 & -2.41 $\pm$ 0.01 & 171.91 $\pm$ 0.04&1013& & &\\ 
         &                     &   \textit{I} &7.14 $\pm$ 0.01 & 6.88 $\pm$ 0.01 & -1.89 $\pm$ 0.01 & 171.87 $\pm$ 0.04&1287& & &\\ 
Vela1-95& 2011-12-22   &    &    &    &       &                  & &  5908.3 $\pm$ 25.1         & 8.25 $\pm$ 0.02 & 1.4 $\pm$ 0.04\\ 
         &                     &   \textit{B} &7.58 $\pm$ 0.14 & 7.37 $\pm$ 0.14 & -1.78 $\pm$ 0.14 & 172.53 $\pm$ 0.18&304& & &\\ 
         &                     &   \textit{V} &8.08 $\pm$ 0.04 & 7.75 $\pm$ 0.04 & -2.3 $\pm$ 0.04 & 172.3 $\pm$ 0.08&739& & &\\ 
         &                     &   \textit{R} &7.91 $\pm$ 0.01 & 7.55 $\pm$ 0.01 & -2.38 $\pm$ 0.01 & 172.06 $\pm$ 0.05&930& & &\\ 
         &                     &   \textit{I} &7.11 $\pm$ 0.01 & 6.85 $\pm$ 0.01 & -1.91 $\pm$ 0.01 & 171.76 $\pm$ 0.05&1175& & &\\ 
Vela1-95& 2013-12-03   &    &    &    &       &                  & &  5913.5 $\pm$ 17.4         & 8.29 $\pm$ 0.01 & 1.38 $\pm$ 0.02\\ 
         &                     &   \textit{B} &7.59 $\pm$ 0.15 & 7.41 $\pm$ 0.15 & -1.65 $\pm$ 0.17 & 173.13 $\pm$ 0.18&378& & &\\ 
         &                     &   \textit{V} &8.13 $\pm$ 0.04 & 7.83 $\pm$ 0.04 & -2.2 $\pm$ 0.04 & 172.78 $\pm$ 0.08&911& & &\\ 
         &                     &   \textit{R} &7.94 $\pm$ 0.01 & 7.59 $\pm$ 0.01 & -2.32 $\pm$ 0.01 & 172.32 $\pm$ 0.05&1157& & &\\ 
         &                     &   \textit{I} &7.17 $\pm$ 0.01 & 6.93 $\pm$ 0.01 & -1.85 $\pm$ 0.01 & 172.1 $\pm$ 0.05&1486& & &\\ 
Vela1-95& 2013-12-05   &    &    &    &       &                  & &  5869.6 $\pm$ 23.3         & 8.31 $\pm$ 0.01 & 1.36 $\pm$ 0.03\\ 
         &                     &   \textit{B} &7.63 $\pm$ 0.14 & 7.43 $\pm$ 0.14 & -1.72 $\pm$ 0.17 & 172.83 $\pm$ 0.17&301& & &\\ 
         &                     &   \textit{V} &8.17 $\pm$ 0.04 & 7.85 $\pm$ 0.04 & -2.27 $\pm$ 0.04 & 172.56 $\pm$ 0.08&740& & &\\ 
         &                     &   \textit{R} &7.93 $\pm$ 0.01 & 7.58 $\pm$ 0.01 & -2.35 $\pm$ 0.01 & 172.19 $\pm$ 0.05&951& & &\\ 
         &                     &   \textit{I} &7.17 $\pm$ 0.01 & 6.92 $\pm$ 0.01 & -1.85 $\pm$ 0.01 & 172.1 $\pm$ 0.04&1240& & &\\ 
Vela1-95& 2013-12-31   &    &    &    &       &                  & &  5848.3 $\pm$ 21.6         & 8.28 $\pm$ 0.01 & 1.34 $\pm$ 0.03\\ 
         &                     &   \textit{B} &7.68 $\pm$ 0.15 & 7.51 $\pm$ 0.15 & -1.61 $\pm$ 0.15 & 173.19 $\pm$ 0.18&290& & &\\ 
         &                     &   \textit{V} &8.15 $\pm$ 0.04 & 7.86 $\pm$ 0.04 & -2.16 $\pm$ 0.04 & 172.89 $\pm$ 0.08&695& & &\\ 
         &                     &   \textit{R} &7.91 $\pm$ 0.01 & 7.58 $\pm$ 0.01 & -2.28 $\pm$ 0.01 & 172.42 $\pm$ 0.05&880& & &\\ 
         &                     &   \textit{I} &7.14 $\pm$ 0.01 & 6.91 $\pm$ 0.01 & -1.81 $\pm$ 0.01 & 172.22 $\pm$ 0.05&1127& & &\\ 
Vela1-95& 2014-01-06   &    &    &    &       &                  & &  5881.1 $\pm$ 19.3         & 8.32 $\pm$ 0.01 & 1.35 $\pm$ 0.03\\ 
         &                     &   \textit{B} &7.63 $\pm$ 0.12 & 7.45 $\pm$ 0.12 & -1.67 $\pm$ 0.13 & 172.86 $\pm$ 0.14&352& & &\\ 
         &                     &   \textit{V} &8.19 $\pm$ 0.03 & 7.88 $\pm$ 0.03 & -2.25 $\pm$ 0.03 & 172.59 $\pm$ 0.06&874& & &\\ 
         &                     &   \textit{R} &7.95 $\pm$ 0.01 & 7.61 $\pm$ 0.01 & -2.32 $\pm$ 0.01 & 172.35 $\pm$ 0.04&1123& & &\\ 
         &                     &   \textit{I} &7.19 $\pm$ 0.01 & 6.95 $\pm$ 0.01 & -1.85 $\pm$ 0.01 & 172.11 $\pm$ 0.04&1467& & &\\ 
Vela1-95& 2014-12-22   &    &    &    &       &                  & &  5831.1 $\pm$ 31.7         & 8.26 $\pm$ 0.02 & 1.34 $\pm$ 0.04\\ 
         &                     &   \textit{B} &7.57 $\pm$ 0.18 & 7.36 $\pm$ 0.18 & -1.78 $\pm$ 0.18 & 172.5 $\pm$ 0.21&244& & &\\ 
         &                     &   \textit{V} &8.18 $\pm$ 0.05 & 7.82 $\pm$ 0.05 & -2.39 $\pm$ 0.05 & 171.97 $\pm$ 0.1&585& & &\\ 
         &                     &   \textit{R} &7.87 $\pm$ 0.01 & 7.49 $\pm$ 0.01 & -2.43 $\pm$ 0.01 & 171.77 $\pm$ 0.06&742& & &\\ 
         &                     &   \textit{I} &7.12 $\pm$ 0.01 & 6.86 $\pm$ 0.01 & -1.89 $\pm$ 0.01 & 171.87 $\pm$ 0.05&960& & &\\ 
Vela1-95& 2015-01-17   &    &    &    &       &                  & &  5695.0 $\pm$ 83.7         & 8.23 $\pm$ 0.04 & 1.21 $\pm$ 0.09\\
         &                     &   \textit{B} &7.72 $\pm$ 0.24 & 7.51 $\pm$ 0.24 & -1.77 $\pm$ 0.26 & 172.63 $\pm$ 0.28&171& & &\\ 
         &                     &   \textit{V} &8.11 $\pm$ 0.06 & 7.76 $\pm$ 0.06 & -2.37 $\pm$ 0.07 & 172.02 $\pm$ 0.13&415& & &\\ 
         &                     &   \textit{R} &7.81 $\pm$ 0.02 & 7.43 $\pm$ 0.02 & -2.42 $\pm$ 0.02 & 171.8 $\pm$ 0.08&526& & &\\ 
         &                     &   \textit{I} &7.03 $\pm$ 0.01 & 6.76 $\pm$ 0.01 & -1.91 $\pm$ 0.01 & 171.69 $\pm$ 0.08&678& & &\\ 
Vela1-95& 2015-05-30   &    &    &    &       &                  & &  5866.9 $\pm$ 40.9         & 8.18 $\pm$ 0.02 & 1.3 $\pm$ 0.05\\ 
         &                     &   \textit{B} &7.52 $\pm$ 0.25 & 7.16 $\pm$ 0.25 & -2.3 $\pm$ 0.25 & 170.57 $\pm$ 0.31&168& & &\\ 
         &                     &   \textit{V} &8.05 $\pm$ 0.06 & 7.64 $\pm$ 0.06 & -2.53 $\pm$ 0.06 & 171.38 $\pm$ 0.14&422& & &\\ 
         &                     &   \textit{R} &7.84 $\pm$ 0.02 & 7.41 $\pm$ 0.02 & -2.57 $\pm$ 0.02 & 171.21 $\pm$ 0.08&539& & &\\ 
         &                     &   \textit{I} &7.09 $\pm$ 0.01 & 6.8 $\pm$ 0.01 & -2.02 $\pm$ 0.01 & 171.3 $\pm$ 0.08&695& & &\\ 
Vela1-95& 2015-12-29   &    &    &    &       &                  & &  5692.1 $\pm$ 30.6         & 8.34 $\pm$ 0.02 & 1.22 $\pm$ 0.03\\ 
         &                     &   \textit{B} &7.9 $\pm$ 0.18 & 7.72 $\pm$ 0.18 & -1.69 $\pm$ 0.18 & 173.17 $\pm$ 0.2&260& & &\\ 
         &                     &   \textit{V} &8.21 $\pm$ 0.04 & 7.86 $\pm$ 0.04 & -2.37 $\pm$ 0.04 & 172.23 $\pm$ 0.09&673& & &\\ 
         &                     &   \textit{R} &7.91 $\pm$ 0.01 & 7.52 $\pm$ 0.01 & -2.45 $\pm$ 0.01 & 171.82 $\pm$ 0.05&879& & &\\ 
         &                     &   \textit{I} &7.11 $\pm$ 0.01 & 6.86 $\pm$ 0.01 & -1.89 $\pm$ 0.01 & 171.86 $\pm$ 0.05&1184& & &\\ 
BD-144922& 2011-10-02   &    &    &    &       &                  & &  5429.6 $\pm$ 21.0         & 6.14 $\pm$ 0.01 & 1.33 $\pm$ 0.03\\ 
         &                     &   \textit{B} &5.82 $\pm$ 0.04 & -1.06 $\pm$ 0.04 & 5.72 $\pm$ 0.04 & 50.01 $\pm$ 0.11&667& & &\\ 
         &                     &   \textit{V} &6.09 $\pm$ 0.02 & -0.99 $\pm$ 0.02 & 6.01 $\pm$ 0.02 & 50.0 $\pm$ 0.07&1133& & &\\ 
         &                     &   \textit{R} &5.8 $\pm$ 0.01 & -0.88 $\pm$ 0.01 & 5.73 $\pm$ 0.01 & 49.85 $\pm$ 0.06&1180& & &\\ 
         &                     &   \textit{I} &4.96 $\pm$ 0.01 & -0.8 $\pm$ 0.01 & 4.89 $\pm$ 0.01 & 49.29 $\pm$ 0.07&1096& & &\\ 
BD-144922& 2013-03-06   &    &    &    &       &                  & &  5470.4 $\pm$ 18.6         & 6.14 $\pm$ 0.01 & 1.32 $\pm$ 0.03\\ 
         &                     &   \textit{B} &5.79 $\pm$ 0.03 & -0.95 $\pm$ 0.03 & 5.71 $\pm$ 0.03 & 49.5 $\pm$ 0.09&812& & &\\ 
         &                     &   \textit{V} &6.1 $\pm$ 0.02 & -0.93 $\pm$ 0.02 & 6.03 $\pm$ 0.02 & 49.66 $\pm$ 0.06&1296& & &\\ 
         &                     &   \textit{R} &5.83 $\pm$ 0.01 & -0.85 $\pm$ 0.01 & 5.77 $\pm$ 0.01 & 49.58 $\pm$ 0.05&1302& & &\\ 
         &                     &   \textit{I} &5.02 $\pm$ 0.01 & -0.79 $\pm$ 0.01 & 4.96 $\pm$ 0.01 & 49.18 $\pm$ 0.07&1126& & &\\ 
HDE316232& 2010-07-14   &    &    &    &       &                  & &  5591.1 $\pm$ 18.3         & 5.02 $\pm$ 0.01 & 1.22 $\pm$ 0.03\\
         &                     &   \textit{B} &4.68 $\pm$ 0.02 & 4.64 $\pm$ 0.02 & 0.62 $\pm$ 0.02 & 3.61 $\pm$ 0.09&1063& & &\\ 
         &                     &   \textit{V} &4.93 $\pm$ 0.01 & 4.9 $\pm$ 0.01 & 0.55 $\pm$ 0.01 & 3.51 $\pm$ 0.07&1529& & &\\ 
         &                     &   \textit{R} &4.77 $\pm$ 0.01 & 4.75 $\pm$ 0.01 & 0.51 $\pm$ 0.01 & 3.37 $\pm$ 0.06&1499& & &\\ 
         &                     &   \textit{I} &4.21 $\pm$ 0.01 & 4.18 $\pm$ 0.01 & 0.56 $\pm$ 0.01 & 3.53 $\pm$ 0.08&1272& & &\\ 
Hiltner652& 2010-07-20   &    &    &    &       &                  & &  5770.2 $\pm$ 18.3         & 6.46 $\pm$ 0.01 & 1.2 $\pm$ 0.03\\ 
         &                     &   \textit{B} &5.94 $\pm$ 0.04 & 5.94 $\pm$ 0.04 & -0.08 $\pm$ 0.04 & 179.32 $\pm$ 0.11&691& & &\\ 
         &                     &   \textit{V} &6.37 $\pm$ 0.02 & 6.37 $\pm$ 0.02 & -0.21 $\pm$ 0.02 & 179.39 $\pm$ 0.07&1083& & &\\ 
         &                     &   \textit{R} &6.21 $\pm$ 0.01 & 6.21 $\pm$ 0.01 & -0.23 $\pm$ 0.01 & 179.3 $\pm$ 0.06&1106& & &\\ 
         &                     &   \textit{I} &5.6 $\pm$ 0.01 & 5.6 $\pm$ 0.01 & -0.05 $\pm$ 0.01 & 179.43 $\pm$ 0.07&978& & &\\ 
Hiltner652& 2011-07-20   &    &    &    &       &                  & &  5807.4 $\pm$ 31.3         & 6.48 $\pm$ 0.02 & 1.24 $\pm$ 0.05\\ 
         &                     &   \textit{B} &5.95 $\pm$ 0.06 & 5.95 $\pm$ 0.06 & 0.03 $\pm$ 0.05 & 179.85 $\pm$ 0.16&477& & &\\ 
         &                     &   \textit{V} &6.36 $\pm$ 0.03 & 6.36 $\pm$ 0.03 & -0.09 $\pm$ 0.03 & 179.94 $\pm$ 0.1&772& & &\\ 
         &                     &   \textit{R} &6.24 $\pm$ 0.01 & 6.24 $\pm$ 0.01 & -0.12 $\pm$ 0.01 & 179.81 $\pm$ 0.08&797& & &\\ 
         &                     &   \textit{I} &5.62 $\pm$ 0.01 & 5.62 $\pm$ 0.01 & 0.05 $\pm$ 0.01 & 179.93 $\pm$ 0.1&714& & &\\ 
Hiltner652& 2011-07-22   &    &    &    &       &                  & &  5816.4 $\pm$ 26.7         & 6.51 $\pm$ 0.01 & 1.15 $\pm$ 0.04\\ 
         &                     &   \textit{B} &5.97 $\pm$ 0.05 & 5.97 $\pm$ 0.05 & 0.04 $\pm$ 0.05 & 0.09 $\pm$ 0.13&529& & &\\ 
         &                     &   \textit{V} &6.42 $\pm$ 0.02 & 6.42 $\pm$ 0.02 & -0.13 $\pm$ 0.02 & 179.69 $\pm$ 0.09&831& & &\\ 
         &                     &   \textit{R} &6.29 $\pm$ 0.01 & 6.29 $\pm$ 0.01 & -0.13 $\pm$ 0.01 & 179.84 $\pm$ 0.07&839& & &\\ 
         &                     &   \textit{I} &5.71 $\pm$ 0.01 & 5.71 $\pm$ 0.01 & 0.08 $\pm$ 0.01 & 0.07 $\pm$ 0.09&723& & &\\ 
Hiltner652& 2015-02-26   &    &    &    &       &                  & &  5788.8 $\pm$ 30.8         & 6.49 $\pm$ 0.02 & 1.16 $\pm$ 0.05\\ 
         &                     &   \textit{B} &5.96 $\pm$ 0.06 & 5.96 $\pm$ 0.06 & 0.02 $\pm$ 0.06 & 179.94 $\pm$ 0.16&436& & &\\ 
         &                     &   \textit{V} &6.41 $\pm$ 0.03 & 6.41 $\pm$ 0.03 & -0.17 $\pm$ 0.03 & 179.55 $\pm$ 0.1&715& & &\\ 
         &                     &   \textit{R} &6.24 $\pm$ 0.01 & 6.23 $\pm$ 0.01 & -0.2 $\pm$ 0.01 & 179.49 $\pm$ 0.08&740& & &\\ 
         &                     &   \textit{I} &5.65 $\pm$ 0.01 & 5.65 $\pm$ 0.01 & -0.05 $\pm$ 0.01 & 179.41 $\pm$ 0.1&675& & &\\ 
Hiltner652& 2015-04-12   &    &    &    &       &                  & &  5795.0 $\pm$ 23.8         & 6.46 $\pm$ 0.01 & 1.15 $\pm$ 0.04\\ 
         &                     &   \textit{B} &5.95 $\pm$ 0.05 & 5.95 $\pm$ 0.05 & -0.11 $\pm$ 0.05 & 179.18 $\pm$ 0.14&516& & &\\ 
         &                     &   \textit{V} &6.36 $\pm$ 0.02 & 6.36 $\pm$ 0.02 & -0.23 $\pm$ 0.02 & 179.29 $\pm$ 0.09&865& & &\\ 
         &                     &   \textit{R} &6.22 $\pm$ 0.01 & 6.21 $\pm$ 0.01 & -0.26 $\pm$ 0.01 & 179.21 $\pm$ 0.07&900& & &\\ 
         &                     &   \textit{I} &5.66 $\pm$ 0.01 & 5.66 $\pm$ 0.01 & -0.06 $\pm$ 0.01 & 179.31 $\pm$ 0.09&794& & &\\ 
Hiltner652& 2015-04-12   &    &    &    &       &                  & &  5759.0 $\pm$ 24.8         & 6.51 $\pm$ 0.01 & 1.16 $\pm$ 0.04\\ 
         &                     &   \textit{B} &6.01 $\pm$ 0.06 & 6.0 $\pm$ 0.06 & -0.09 $\pm$ 0.06 & 179.39 $\pm$ 0.17&569& & &\\ 
         &                     &   \textit{V} &6.41 $\pm$ 0.03 & 6.41 $\pm$ 0.03 & -0.24 $\pm$ 0.03 & 179.26 $\pm$ 0.11&939& & &\\ 
         &                     &   \textit{R} &6.26 $\pm$ 0.01 & 6.26 $\pm$ 0.01 & -0.22 $\pm$ 0.01 & 179.38 $\pm$ 0.09&968& & &\\ 
         &                     &   \textit{I} &5.66 $\pm$ 0.01 & 5.66 $\pm$ 0.01 & -0.06 $\pm$ 0.01 & 179.37 $\pm$ 0.11&843& & &\\ 
Hiltner652& 2015-05-09   &    &    &    &       &                  & &  5763.9 $\pm$ 32.9         & 6.4 $\pm$ 0.02 & 1.2 $\pm$ 0.05\\ 
         &                     &   \textit{B} &5.88 $\pm$ 0.05 & 5.88 $\pm$ 0.05 & -0.07 $\pm$ 0.05 & 179.46 $\pm$ 0.15&479& & &\\ 
         &                     &   \textit{V} &6.31 $\pm$ 0.03 & 6.31 $\pm$ 0.03 & -0.24 $\pm$ 0.03 & 179.23 $\pm$ 0.1&756& & &\\ 
         &                     &   \textit{R} &6.15 $\pm$ 0.01 & 6.15 $\pm$ 0.01 & -0.26 $\pm$ 0.01 & 179.21 $\pm$ 0.08&773& & &\\ 
         &                     &   \textit{I} &5.52 $\pm$ 0.01 & 5.52 $\pm$ 0.01 & -0.11 $\pm$ 0.01 & 179.07 $\pm$ 0.1&709& & &\\ 
Hiltner652& 2015-06-10   &    &    &    &       &                  & &  5726.7 $\pm$ 25.8         & 6.42 $\pm$ 0.01 & 1.16 $\pm$ 0.04\\ 
         &                     &   \textit{B} &5.94 $\pm$ 0.04 & 5.94 $\pm$ 0.04 & -0.11 $\pm$ 0.04 & 179.26 $\pm$ 0.12&600& & &\\ 
         &                     &   \textit{V} &6.34 $\pm$ 0.02 & 6.33 $\pm$ 0.02 & -0.25 $\pm$ 0.02 & 179.25 $\pm$ 0.08&933& & &\\ 
         &                     &   \textit{R} &6.17 $\pm$ 0.01 & 6.16 $\pm$ 0.01 & -0.27 $\pm$ 0.01 & 179.05 $\pm$ 0.07&949& & &\\ 
         &                     &   \textit{I} &5.53 $\pm$ 0.01 & 5.53 $\pm$ 0.01 & -0.1 $\pm$ 0.01 & 179.14 $\pm$ 0.08&831& & &\\ 
NGC2024& 2010-12-31   &    &    &    &       &                  & &  6355.6 $\pm$ 10.1         & 9.87 $\pm$ 0.01 & 1.31 $\pm$ 0.02\\ 
         &                     &   \textit{B} &8.53 $\pm$ 0.1 & 0.74 $\pm$ 0.1 & -8.5 $\pm$ 0.1 & 136.8 $\pm$ 0.1&449& & &\\ 
         &                     &   \textit{V} &9.55 $\pm$ 0.03 & 0.25 $\pm$ 0.03 & -9.54 $\pm$ 0.03 & 136.33 $\pm$ 0.05&1008& & &\\ 
         &                     &   \textit{R} &9.67 $\pm$ 0.01 & 0.09 $\pm$ 0.01 & -9.67 $\pm$ 0.01 & 136.19 $\pm$ 0.03&1314& & &\\ 
         &                     &   \textit{I} &9.02 $\pm$ 0.0 & 0.49 $\pm$ 0.0 & -9.0 $\pm$ 0.0 & 136.18 $\pm$ 0.02&1832& & &\\ 
NGC2024& 2011-12-30   &    &    &    &       &                  & &  6337.7 $\pm$ 11.7         & 9.78 $\pm$ 0.01 & 1.26 $\pm$ 0.02\\ 
         &                     &   \textit{B} &8.56 $\pm$ 0.12 & 0.63 $\pm$ 0.11 & -8.54 $\pm$ 0.12 & 136.62 $\pm$ 0.12&408& & &\\ 
         &                     &   \textit{V} &9.46 $\pm$ 0.03 & 0.13 $\pm$ 0.03 & -9.46 $\pm$ 0.03 & 136.0 $\pm$ 0.05&927& & &\\ 
         &                     &   \textit{R} &9.59 $\pm$ 0.01 & -0.02 $\pm$ 0.01 & -9.59 $\pm$ 0.01 & 135.91 $\pm$ 0.03&1214& & &\\ 
         &                     &   \textit{I} &8.95 $\pm$ 0.01 & 0.34 $\pm$ 0.0 & -8.95 $\pm$ 0.01 & 135.73 $\pm$ 0.02&1692& & &\\ 
NGC2024& 2012-11-14   &    &    &    &       &                  & &  6308.6 $\pm$ 12.4         & 9.95 $\pm$ 0.01 & 1.26 $\pm$ 0.02\\ 
         &                     &   \textit{B} &8.7 $\pm$ 0.12 & 0.74 $\pm$ 0.11 & -8.67 $\pm$ 0.12 & 137.0 $\pm$ 0.11&408& & &\\ 
         &                     &   \textit{V} &9.64 $\pm$ 0.03 & 0.27 $\pm$ 0.03 & -9.63 $\pm$ 0.03 & 136.39 $\pm$ 0.05&911& & &\\ 
         &                     &   \textit{R} &9.75 $\pm$ 0.01 & 0.18 $\pm$ 0.01 & -9.75 $\pm$ 0.01 & 136.47 $\pm$ 0.03&1193& & &\\ 
         &                     &   \textit{I} &9.08 $\pm$ 0.01 & 0.56 $\pm$ 0.01 & -9.06 $\pm$ 0.01 & 136.4 $\pm$ 0.02&1662& & &\\ 
NGC2024& 2015-05-11   &    &    &    &       &                  & &  6366.5 $\pm$ 13.0         & 10.0 $\pm$ 0.01 & 1.27 $\pm$ 0.02\\ 
         &                     &   \textit{B} &8.69 $\pm$ 0.09 & 0.37 $\pm$ 0.11 & -8.68 $\pm$ 0.09 & 135.74 $\pm$ 0.12&364& & &\\ 
         &                     &   \textit{V} &9.64 $\pm$ 0.03 & -0.04 $\pm$ 0.04 & -9.64 $\pm$ 0.03 & 135.31 $\pm$ 0.06&735& & &\\ 
         &                     &   \textit{R} &9.86 $\pm$ 0.01 & -0.11 $\pm$ 0.01 & -9.85 $\pm$ 0.01 & 135.48 $\pm$ 0.04&845& & &\\ 
         &                     &   \textit{I} &9.19 $\pm$ 0.01 & 0.3 $\pm$ 0.01 & -9.18 $\pm$ 0.01 & 135.59 $\pm$ 0.04&1016& & &\\ 
NGC2024& 2015-09-25   &    &    &    &       &                  & &  6310.4 $\pm$ 12.9         & 9.81 $\pm$ 0.01 & 1.27 $\pm$ 0.02\\ 
         &                     &   \textit{B} &8.57 $\pm$ 0.11 & 0.52 $\pm$ 0.11 & -8.55 $\pm$ 0.11 & 136.22 $\pm$ 0.11&427& & &\\ 
         &                     &   \textit{V} &9.51 $\pm$ 0.03 & 0.04 $\pm$ 0.03 & -9.51 $\pm$ 0.03 & 135.68 $\pm$ 0.05&944& & &\\ 
         &                     &   \textit{R} &9.62 $\pm$ 0.01 & -0.1 $\pm$ 0.01 & -9.62 $\pm$ 0.01 & 135.64 $\pm$ 0.03&1230& & &\\ 
         &                     &   \textit{I} &8.95 $\pm$ 0.0 & 0.3 $\pm$ 0.01 & -8.94 $\pm$ 0.0 & 135.58 $\pm$ 0.02&1703& & &\\ 
NGC2024& 2015-10-14   &    &    &    &       &                  & &  6363.9 $\pm$ 14.9         & 9.81 $\pm$ 0.01 & 1.3 $\pm$ 0.02\\ 
         &                     &   \textit{B} &8.47 $\pm$ 0.15 & 0.41 $\pm$ 0.16 & -8.46 $\pm$ 0.15 & 135.91 $\pm$ 0.15&306& & &\\ 
         &                     &   \textit{V} &9.49 $\pm$ 0.04 & -0.0 $\pm$ 0.04 & -9.49 $\pm$ 0.04 & 135.61 $\pm$ 0.07&701& & &\\ 
         &                     &   \textit{R} &9.62 $\pm$ 0.01 & -0.14 $\pm$ 0.01 & -9.62 $\pm$ 0.01 & 135.53 $\pm$ 0.04&927& & &\\ 
         &                     &   \textit{I} &8.99 $\pm$ 0.01 & 0.31 $\pm$ 0.01 & -8.98 $\pm$ 0.01 & 135.6 $\pm$ 0.03&1327& & &\\ 
NGC2024& 2015-10-14   &    &    &    &       &                  & &  6339.6 $\pm$ 13.7         & 9.81 $\pm$ 0.01 & 1.25 $\pm$ 0.02\\ 
         &                     &   \textit{B} &8.58 $\pm$ 0.21 & 0.57 $\pm$ 0.21 & -8.56 $\pm$ 0.21 & 136.29 $\pm$ 0.2&309& & &\\ 
         &                     &   \textit{V} &9.5 $\pm$ 0.06 & 0.05 $\pm$ 0.06 & -9.5 $\pm$ 0.06 & 135.73 $\pm$ 0.09&710& & &\\ 
         &                     &   \textit{R} &9.62 $\pm$ 0.01 & -0.09 $\pm$ 0.01 & -9.62 $\pm$ 0.01 & 135.7 $\pm$ 0.05&943& & &\\ 
         &                     &   \textit{I} &8.98 $\pm$ 0.01 & 0.37 $\pm$ 0.01 & -8.97 $\pm$ 0.01 & 135.82 $\pm$ 0.04&1358& & &\\ 
BD-125133& 2015-09-08   &    &    &    &       &                  & &  5049.5 $\pm$ 35.5         & 4.37 $\pm$ 0.01 & 1.17 $\pm$ 0.04\\ 
         &                     &   \textit{B} &4.22 $\pm$ 0.03 & 1.69 $\pm$ 0.03 & -3.86 $\pm$ 0.03 & 146.54 $\pm$ 0.12&845& & &\\ 
         &                     &   \textit{V} &4.27 $\pm$ 0.02 & 1.57 $\pm$ 0.02 & -3.97 $\pm$ 0.02 & 145.88 $\pm$ 0.09&1252& & &\\ 
         &                     &   \textit{R} &4.0 $\pm$ 0.01 & 1.41 $\pm$ 0.01 & -3.74 $\pm$ 0.01 & 145.62 $\pm$ 0.08&1213& & &\\ 
         &                     &   \textit{I} &3.35 $\pm$ 0.01 & 1.22 $\pm$ 0.01 & -3.12 $\pm$ 0.01 & 145.28 $\pm$ 0.11&1049& & &\\ 
\end{longtable}
The polarization angle, $\theta$, in individual bands is the average polarization angle from 3980-4920$\AA$ (\textit{B} band), 5070-5950$\AA$ (\textit{V}), 5890-5890$\AA$ (\textit{R}) and 7310-8810$\AA$ (\textit{I}). The signal-to-noise ratio, S/N, in individual bands is the average SNR between two 50$\AA$ bins, from 4250-4350$\AA$ (\textit{B}), 5450-5550$\AA$ (\textit{V}), 6250-6350$\AA$ (\textit{R}) and 7650-7850$\AA$ (\textit{I}). 
}

\bsp	
\label{lastpage}
\end{document}